\documentclass{jfm}
\usepackage{graphicx,epsfig}
\usepackage{natbib}
\usepackage{amssymb,amsbsy,upmath}
\usepackage{amsmath}
\usepackage{lscape}
\usepackage{pstricks}

\newcommand{\al}{\alpha}
\newcommand{\K}{\mathrm{K}}
\newcommand{\E}{\mathrm{E}}

\title[Transcritical flow of a stratified fluid over topography]{Transcritical flow of a stratified fluid over topography:
analysis of the forced Gardner equation}

\author[A. M. Kamchatnov et al.]{ A. M. KAMCHATNOV$^{1}$,  Y.-H. KUO$^{2}$,  T.-C. LIN$^{2}$,  T.-L. HORNG$^{3}$,  \\ S.-C. GOU$^{4}$,   R. CLIFT$^{5}$,  G. A. EL$^{5}$,  R. H. J. GRIMSHAW$^{5}$}

\affiliation{$^1$Institute of Spectroscopy, Russian Academy of Sciences, \\
Troitsk, Moscow, 142190, Russia, \\ $^{2}$ Department of Mathematics, National Taiwan University, Taipei, Taiwan, \\
$^{3}$ Department of Applied Mathematics, Feng Chia University, Taichung 40724, Taiwan, \\
$^{4}$ Department of Physics, National Changhua University of Education, \\ Changhua 50058, Taiwan, \\
$^5$Department of Mathematical Sciences, Loughborough University, \\
Loughborough LE11 3TU, UK }

\date{}
\begin{document}

\maketitle

\begin{abstract}
Transcritical flow of a stratified fluid past a broad localised topographic obstacle is studied analytically
in the framework of the forced extended Korteweg--de Vries (eKdV), or Gardner, equation.
We consider both possible signs for the cubic nonlinear
term in the Gardner equation corresponding to different  fluid density stratification profiles.
We identify the range of the input parameters: the oncoming flow speed (the Froude number) and the
topographic amplitude, for which the obstacle supports a stationary localised hydraulic transition
from the subcritical flow upstream to the supercritical flow downstream.
Such a localised transcritical flow is resolved back into the equilibrium flow state away from the obstacle
with the aid of  unsteady coherent nonlinear wave structures propagating upstream and downstream.
Along with the regular, cnoidal undular bores  occurring in the analogous problem for the single-layer flow modeled by the forced KdV equation, the transcritical internal wave flows support a diverse family of upstream and downstream wave structures,
including solibores, rarefaction waves, reversed  and trigonometric undular bores,
which we describe using the recent development of the nonlinear modulation theory for the (unforced) Gardner equation.
The predictions of the developed analytic construction are confirmed by direct numerical simulations of the forced Gardner
equation for a broad range of input parameters.

\end{abstract}

\section{Introduction}

The problem of weakly nonlinear long wave evolution over localized obstacles is of fundamental importance
in various branches of fluid mechanics and has a number of applications in oceanography and meteorology.
The particular case of one-dimensional flow propagation over a broad ridge is quite remarkable due to the existence of
a generic mechanism of the wave generation when the speed of the flow becomes close to the local speed of linear long waves in
the reference frame of the obstacle. This mechanism is essentially nonlinear (see, e.g.,
\cite{akylas84,cole85}) and its mathematical description requires inclusion of appropriately balanced forcing,
nonlinear and dispersive terms in the corresponding asymptotic model.

The prominent feature of a near-critical flow  past broad obstacles is the  formation of a smooth steady (hydraulic)
transition over the obstacle region  so that the flow in front of the obstacle is subcritical and after the obstacle supercritical (see \cite{baines} and references therein).
The upstream and downstream non-equilibrium states are resolved  back into the equilibrium flow away from the obstacle with the aid of unsteady nonlinear wave trains,  or undular bores. The general mathematical framework for the study of transcritical weakly nonlinear
and weakly dispersive fluid flows was introduced in \cite{gs-1986}.  In this work a combination of the dispersionless
hydraulic solution of the forced Korteweg--de Vries (KdV) equation for the near-field and the solutions
of the KdV--Whitham equations describing the modulations in the undular bores propagating away from the obstacle
were used. This Grimshaw--Smyth (GS) formulation has been widely used in further studies of transcritical shallow-water
flows (see \cite{madsen-2012} and references therein) but has also proved to be relevant to other physical
contexts such as  superfluids, \cite{legk-2009}, and nonlinear optics, \cite{fleischer}, when the ``fluid''
dynamics is governed by the defocusing Gross-Pitaevskii/nonlinear Schr\"odinger equation.

If the forcing amplitude is small, the dynamics of the transcritical shallow-water flow (both for surface and
internal waves) is typically
governed by the forced KdV equation (see the reviews \cite{baines,grimshaw-2001,hm-2006,aosl-2007}).
However, for sufficiently large forcing or for special conditions of  the background density stratification
the description of internal waves generated in stratified fluid flows often
requires the inclusion of higher-order nonlinear terms resulting, for near-critical
flows, in the forced extended KdV (eKdV), or Gardner, equation, which in
non-dimensional variables has the form (see, e.g., \cite{ky-1978,mh-1987})
\begin{equation}\label{1-1}
    -u_t-\Delta u_x+6uu_x-6\al u^2u_x+u_{xxx}+G_x=0.
\end{equation}
Here $u$ denotes an appropriate field variable (e.g., the interfacial displacement in a two-layer fluid),
$\al$ measures the relative value of the  cubic nonlinear term {\it vis-a-vis} the quadratic nonlinear term, $\Delta$ is the deviation of
the incident flow velocity from the long wave phase speed and $G(x)$ represents the localized topographic forcing,
that is we assume that
\begin{equation}\label{1-2}
    G(x)\to0\quad \mathrm{at} \quad x\to\pm\infty;
\end{equation}
$G(x)$ can be considered as vanishing outside the region $|x|\gtrsim l$ and having
its maximal value $G_m$ at $x=0$. We shall be interested here only in the
positive forcing case, $G(x)>0$. In numerical simulations $G(x)$ is often taken as
\begin{equation}\label{1-3}
    G(x)=G_m\exp(-x^2/l^2).
\end{equation}
At the initial moment of time it is assumed that
\begin{equation}\label{1-4}
    u(x,0)=0 ,
\end{equation}
which corresponds to  ``turning on'' the forcing at $t=0$. Thus
the flow satisfies the boundary condition
\begin{equation}\label{1-5}
    u(x,t)\to 0 \quad \mathrm{for} \quad |x|\to\infty.
\end{equation}
The problem (\ref{1-1}), (\ref{1-2}), (\ref{1-4}), (\ref{1-5}) was considered using a number of approaches, but mostly numerically,
in \cite{mh-1987}, \cite{ms-1990} and \cite{gcc-2002}. In \cite{ms-1990} an analytical treatment of the forced Gardner
equation was restricted to the case of small positive values of $\alpha$ in
(\ref{1-1}) which enables one to capture only the small quantitative corrections to the wave
regimes described by the forced KdV equation. In \cite{gcc-2002} the transcritical hydraulic
solution of the forced Gardner equation was constructed for finite values of $\alpha$ but the
unsteady wavetrains generated upstream and downstream were treated only numerically.
This latter study revealed a broad variety of the wave regimes realised in transcritical
internal wave flows, part of which were observed in an earlier work \cite{mh-1987}.
Importantly, some of these numerically obtained regimes for the forced Gardner equation were sharply different from those occurring in the forced KdV dynamics \cite{gs-1986}. In particular, the numerical solutions of the forced  Gardner equation (\ref{1-1}) with $\alpha >0$ presented in \cite{mh-1987} and \cite{gcc-2002} showed that, for a certain range of the problem input parameters $G_m$ and $\Delta$ the upstream structure consists of the combination of a rarefaction wave and a solibore (a dissipationless monotone bore solution of the Gardner equation) rather than the cnoidal undular bore observed in the counterpart KdV setting, when $\alpha = 0$. Other generated patterns occurring in the numerical simulations in \cite{gcc-2002} include solitary wavetrains and irregularly generated waves of variable amplitude which may continually interact. Unlike the forced KdV equation, the forced Gardner equation was shown to support propagating wave structures in subcritical and supercritical flow regimes. In this paper we restrict our study to the transcritical flow  regimes when the
obstacle supports a stationary localised hydraulic transition from the subcritical flow upstream to the supercritical flow downstream. In this case an analytical description of the unsteady waves generated upstream and downstream in the framework of the general GS setting becomes possible in principle. The practical realisation of this possibility, however, had been hindered until recently due to the unavailability of the full modulation (Whitham) theory for the Gardner equation.

The modulation theory for the Gardner equation was developed in the recent study \cite{kamch-12} where the complete classification of the initial step resolution (Riemann) problem was constructed.  The generated structures, along with the KdV type cnoidal undular bores, include nonlinear `trigonometric'
bores, solibores, rarefaction waves and composite solutions representing various combinations of the above structures.  In the current paper, we  utilise the analytical results of \cite{kamch-12} to describe and classify  the  wave patterns occurring in transcritical internal wave flows over broad localized topography.  We consider both signs of the  coefficient of the cubic nonlinear term,  $\alpha$ in (\ref{1-1}) corresponding to different background density stratification conditions (see e.g. \cite{grimshaw-2001}).

The structure of the paper is as follows. In section 2 we present a detailed study of the solutions to the forced Gardner equation  (\ref{1-1}) in the hydraulic (steady dispersionless) approximation. These are needed, in particular, to derive the values of the upstream and downstream hydraulic jumps forming in the transcritical flow regime. In section 3 the full classification of solutions to the initial step resolution (Riemann) problem for the Gardner equation is presented following the recent study by \cite{kamch-12}. The results of sections 2 and 3 are then used in sections 4 and 5 to describe the unsteady wave patterns occurring in the dispersive resolution of the upstream and downstream hydraulic jumps in the transcritical flows governed by the Gardner equation (\ref{1-1}) with a broad forcing for both signs of $\alpha$.   The predictions of the analytical theory developed in sections 3--5 are confirmed by the direct numerical simulations of the problem  (\ref{1-1})--(\ref{1-5}).  In section 6 we present conclusions.

\section{Hydraulic approximation}
In this section we present an analysis of the  solutions to the forced Gardner equation in the hydraulic approximation which is the key to the GS construction employed in this paper. The conditions for the existence of the steady localised transcritical hydraulic solution giving rise to the generation of unsteady wave structures upstream and downstream of the obstacle  were identified in \cite{gcc-2002}. Below we review, detail and extend the results of \cite{gcc-2002} by identifying, within the region of existence of the transcritical hydraulic transition, the further three subregions (for either sign of $\alpha$) corresponding to various unsteady wave regimes occurring away from the obstacle. This is done using the full classification of the solutions to the Riemann step problem for the Gardner equation obtained in \cite{kamch-12}.

\subsection{Subcritical and supercritical flows}

We assume that in non-dimensional units $l\gg1$. It is then natural to expect that the topographic forcing
results in a disturbance  $u$ also with the characteristic length $\sim l$.
Then the dispersion term $u_{xxx}\sim l^{-3}$
is relatively small compared with the other terms $\sim l^{-1}$ and we can describe the motion
by the dispersionless limit of (\ref{1-1})
\begin{equation}\label{1-6}
    -u_t-\Delta u_x+6uu_x-6\al u^2u_x+G_x=0.
\end{equation}
The ``switching on'' the forcing  generates the disturbance in the region $|x|\lesssim l$
of the location of the
topography. One can then expect that some part of the disturbance will leave this region with the group
velocity and, in the limit $t\to\infty$, can be neglected whereas the remaining part
can be described by
a stationary solution of equation (\ref{1-6}), that is
\begin{equation}\label{1-7}
    -\Delta u_x+6uu_x-6\al u^2u_x+G_x=0 \, ,
\end{equation}
with the boundary condition (\ref{1-5}). Approximation (\ref{1-7}) is usually
called the {\it hydraulic approximation}.  Integrating (\ref{1-7}) and using the boundary
condition (\ref{1-5}),
\begin{equation}\label{2-1}
    -\Delta u+3u^2-2\al u^3 + G(x) =0,
\end{equation}
yields the dependence $u(x)$ in an implicit form for a given function $G(x)$. Let us consider
the conditions for the existence of such a solution for all $-\infty<x<\infty$.

Equation (\ref{2-1}) can be written in the form,
\begin{equation}\label{2-2}
    F(u)=\Delta u-3u^2+2\al u^3 =G(x) \,,
\end{equation}
where the polynomial $F(u)$ has three roots, one of which is $u=0$, corresponding to the equilibrium state at $x\to\pm\infty$.
If the other two roots are real,
then, since $G>0$, the solution of (\ref{2-1}) is possible for $F(u)>0$ only, and it can be found
for all $x\in(-\infty,\infty)$ provided the following condition is satisfied:
\begin{equation}\label{2-3}
    G_m\leq F_m,
\end{equation}
where $G_m$ is the amplitude of the forcing and $F_m$ is a local maximum of the function $F(u)$, which exists for
\begin{equation}\label{2-5a}
    \Delta<\frac{3}{2\alpha}\quad\text{for}\quad \alpha>0 \,, \quad \text{or}\quad
    \Delta>\frac{3}{2\alpha}\quad\text{for}\quad \alpha<0 \,.
\end{equation}
Otherwise this maximum disappears and the hydraulic solution is guaranteed to exist
for any value of $G_m$.

It is easy to find that $F(u)$ reaches its maximum value at
\begin{equation}\label{2-4}
    u_m=\frac1{2\al}\left(1-\sqrt{1-\frac{2\al\Delta}3}\right)
\end{equation}
and it is given by the formula
\begin{equation}\label{2-5}
    F_m(\Delta)=F(u_m)=-\frac1{2\al^2}\left[1-\left(1-\frac{2\al\Delta}3\right)^{3/2}\right]
    +\frac{\Delta}{2\al}.
\end{equation}
 For future reference we indicate here the value
\begin{equation}\label{2-4a}
    u_l=\frac1{2\al}\left[1+\sqrt{1-\frac{2\al\Delta}3}\right]  \, ,
\end{equation}
at which $F(u)$ has a local minimum equal to
\begin{equation}\label{2-5b}
    F_l(\Delta)=F(u_l)=-\frac1{2\al^2}\left[1+\left(1-\frac{2\al\Delta}3\right)^{3/2}\right]
    +\frac{\Delta}{2\al}.
\end{equation}

\begin{figure}
\begin{center}
\includegraphics[width=7cm, clip]{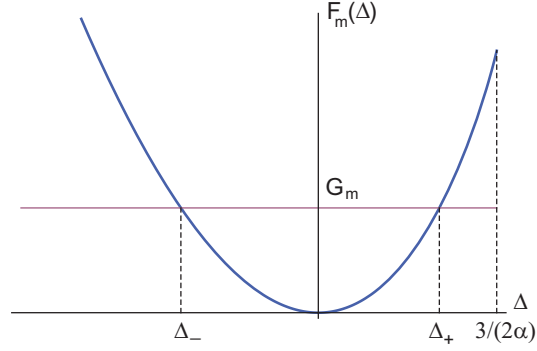}
\caption{Plot of the function $F_m(\Delta)$ defined by Eq.~(\ref{2-5})  with $\alpha>0$.
Subcritical $\Delta<\Delta_-$ and supercritical $\Delta>\Delta_+$ regions correspond to the
values of $\Delta$ when the condition $G_m\leq F_m(\Delta)$ is satisfied.
 }
\end{center}\label{fig1}
\end{figure}

The dependence of $F_m(\Delta)$ is illustrated in Fig.~1 for the case $\alpha>0$.
The condition (\ref{2-3}) is fulfilled in the regions
\begin{equation}\label{2-6}
    -\infty<\Delta<\Delta_- \quad \text{and}\quad \Delta_+<\Delta<\frac3{2\alpha},
\end{equation}
where $\Delta_{\pm}$ are determined by the equation  $F_m(\Delta)=G_m$, that is,
\begin{equation}\label{2-7}
   -\frac1{2\al^2}\left[1-\left(1-\frac{2\al\Delta}3\right)^{3/2}\right]+\frac{\Delta}{2\al} =G_m \, .
\end{equation}
When $\alpha > 0$ then $\Delta_{+} \le 3/2\alpha $ for $G_m \le  1/4\alpha^2 $
and increases monotonically as $G_m $ increases, and
$\Delta_{+} = 3/2\alpha $ for $G_m > 1/4\alpha^2 $, while  when $\alpha < 0$,
$\Delta_{+}$ increases monotonically as $G_m $ increases.  Similarly,
when $\alpha > 0 $ then $\Delta_{-} $  decreases monotonically as $G_m $ increases,
while when $\alpha < 0$, $\Delta_{-} \ge 3/2\alpha $ for $G_m \le  1/4\alpha^2 $
and decreases monotonically as $G_m $ increases, and
$\Delta_{-} = 3/2\alpha $ for $G_m > 1/4\alpha^2 $.
The region $-\infty<\Delta<\Delta_-$ is subcritical and the region
$\Delta_+<\Delta<+\infty$ is supercritical. In both these regions
a stationary global solution exists in the hydraulic
approximation with $u$ continuously depending on $x$ on the entire axis.
These solutions are illustrated in Fig.~2. As we can see, the form of the near-field reproduces qualitatively
the form of the obstacle and the polarity of the disturbance depends on the value of
$\Delta$. It is a localised depression for subcritical $\Delta$ and a localised elevation for supercritical $\Delta$.
The supercritical (subcritical) solution has  a maximum (minimum) $u_m $.

It  readily follows from  (\ref{2-7}) that the dependence of $\Delta_{\pm}$ on $G_m$
can be expressed in an explicit form,
\begin{equation}\label{3-4b2}
    \Delta_{\pm}=\frac1{\alpha}\Phi_{\pm}(\alpha^2G_m) \quad\text{for}\quad\alpha>0\quad\text{or}\quad 
    \Delta_{\pm}=\frac1{\alpha}\Phi_{\mp}(\alpha^2G_m) \quad\text{for}\quad\alpha<0,
\end{equation}
where $\Phi_{\pm}(y)$ are determined as functions inverse to
\begin{equation}\label{3-5a}
    y=\Phi^{-1}(x)=\frac12\left[x-1+\left(1-\frac{2x}3\right)^{3/2}\right].
\end{equation}
They can be represented in the explicit form,
\begin{equation}\label{3-5b}
   \Phi_{\pm}(y)= \frac38\left(1-2\sqrt{1+32y}\,\cos\frac{\theta(y)\pm\pi}3\right),
\end{equation}
where
\begin{equation}\label{3-5c}
    \theta(y)=\arccos\left(\frac{1-80y-128y^2}{(1+32y)^{3/2}}\right),\quad 0\leq y\leq 1/4.
\end{equation}
In the limit of small $|\al|$ we find the series expansion
\begin{equation}\label{3-2}
    F_m(\al,\Delta)\cong \frac{\Delta^2}{12}+\frac{\al\Delta^3}{108}+\frac{\al^2\Delta^4}{432}+\ldots,
\end{equation}
and, correspondingly,
\begin{equation}\label{3-3}
    \begin{split}
    &\Delta_-\cong -\sqrt{12G_m}-\frac{2\al G_m}3+\frac{4\al^2G_m^{3/2}}{9\sqrt{3}}+\ldots,\\
    &\Delta_+\cong \sqrt{12G_m}-\frac{2\al G_m}3-\frac{4\al^2G_m^{3/2}}{9\sqrt{3}}+\ldots,
    \end{split}
\end{equation}
The first term of this expansion corresponds to the forced KdV result of \cite{gs-1986}.

\begin{figure}
\begin{center}
\includegraphics[width=6cm,clip]{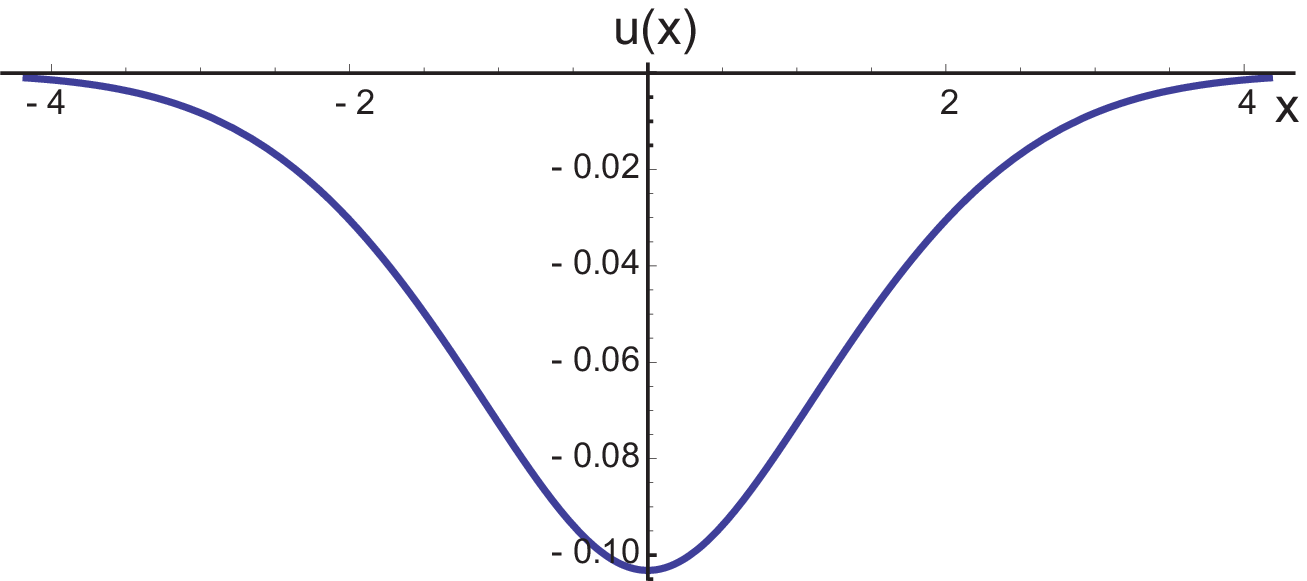} \qquad
\includegraphics[width=6cm,clip]{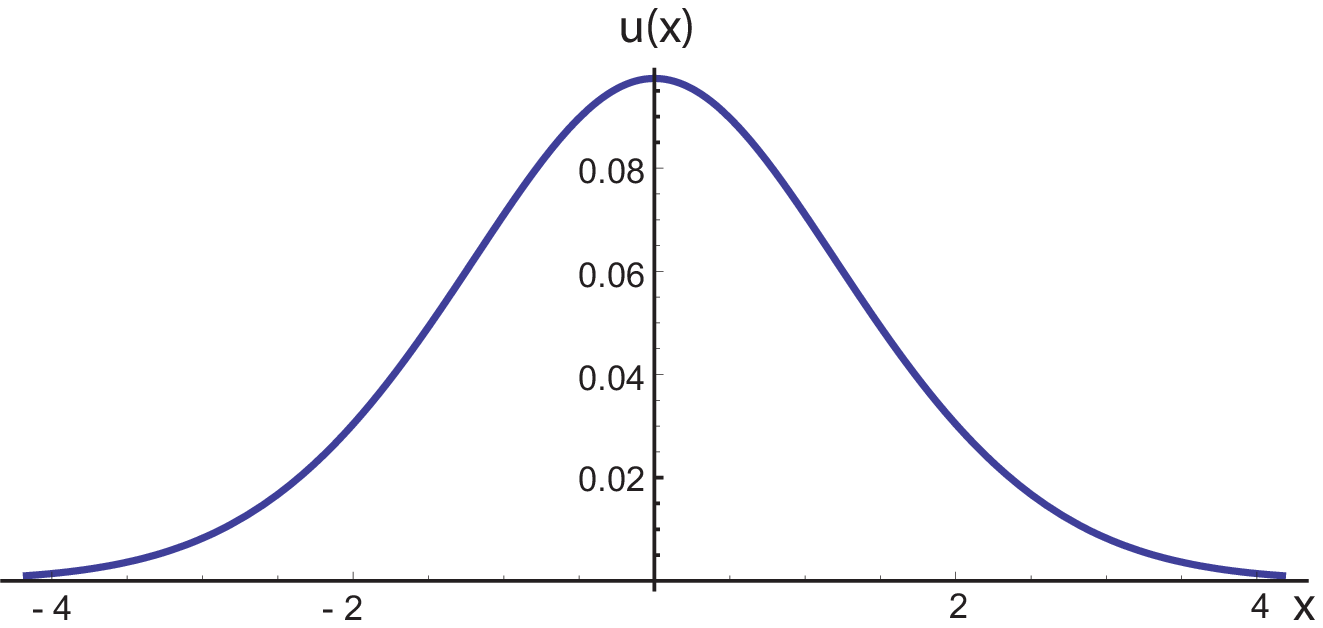}
\caption{The hydraulic solution $u(x)$ for subcritical and supercritical values of $\Delta$:
$\Delta=-1.3$ (left) and $\Delta= 1.3$
(right). For $G_m=0.1$ and $\alpha=1$ the critical values are equal to $\Delta_-=-1.1555$ and
$\Delta_+=1.0177$; the width of the
forcing (\ref{1-3}) equals to $l=2$.
 }
\end{center}\label{fig2}
\end{figure}

\subsection{Transcritical region}

If $\Delta$ is located in the {\it transcritical} region
\begin{equation}\label{2-8}
    \Delta_-<\Delta<\Delta_+,
\end{equation}
then the global smooth steady hydraulic solution defined for all $x \in (-\infty, +\infty)$ and described in section 2.1,
ceases to exist. Nevertheless, in the region  $-l\lesssim x\lesssim l$
of the obstacle's location the hydraulic approximation can  still be applicable, but now, considered globally,
the hydraulic solution no longer satisfies the equilibrium boundary condition (\ref{1-5}). Rather,  instead there
exists a region of the flow parameters
when the hydraulic solution tends to some nonzero constant values,
\begin{equation}\label{3-1}
    u(x)\to\left\{
    \begin{array}{cl}
    u^u,\quad &x\to-\infty,\\
    u^d,\quad &x\to+\infty,
    \end{array}
    \right.
\end{equation}
where the upstream and downstream values $u^{u}$ and $u^d$ respectively are determined by these parameters.
Of course, the limits $x\to\pm\infty$ have here a
formal sense; the limiting values $u^{u,d}$ are practically reached at $x\sim\pm l$ (we remind that $l \gg1$)
and to satisfy the equilibrium boundary conditions
at infinity,  upstream and downstream undular bores must be inserted (see \cite{gs-1986}).

Since the boundary condition (\ref{1-5}) is no longer applicable in the transcritical region,
one should use instead of  (\ref{2-1}) the general integral of Eq.~(\ref{1-7}),
\begin{equation}\label{3-4}
    F(u) + C = G(x) \,,
\end{equation}
where $C$ is an integration constant, and  can be found in the following way.
Let the maximal value $G_m$ of $G(x)$ be located at $x=0$, so that $G_x(0)=0$. We look for the
{\it transcritical} solution with $u_x(0)\neq 0$; hence Eq.~(\ref{1-7}) yields
\begin{equation}\label{3-5}
    u_m\equiv u(0)=\frac1{2\al}\left(1-\sqrt{1-\frac{2\al\Delta}3}\right)
\end{equation}
which coincides with (\ref{2-4}). Hence, $u(x)$ is determined by the equation
\begin{equation}\label{3-6a}
    F(u)-G(x)=F_m(\Delta)- G_m \,,
\end{equation}
\begin{equation}\label{3-6}
\hbox{or} \quad     2\al u^3 - 3u^2 + \Delta u - G(x)=
   - \frac1{2\al^2} \left[1-\left(1-\frac{2\al\Delta}3\right)^{3/2}\right]
   + \frac{\Delta}{2\al} - G_m.
\end{equation}
In the formal limit $x\to\pm\infty$ (actually, for $|x|\sim l$) we have $G(x)\to0$ and
the limiting values $u^{u,d}$ of $u$ are the roots of the equation
\begin{equation}\label{3-7}
    2\al u^3-3u^2+\Delta u=-\frac1{2\al^2}\left[1-\left(1-\frac{2\al\Delta}3\right)^{3/2}\right]
    + \frac{\Delta}{2\al}-G_m.
\end{equation}
Although this equation has three roots, only two of them neighboring $u_m$ have
physical meaning since they correspond
to the limiting values of a continuous solution of Eq.~(\ref{3-6}) in the whole
range of the values of $x$ in the interval $[-l, l]$.
The upstream and downstream values $u^u$ and $u^d$ are ordered according to
\begin{equation}\label{4-1}
    u^d<u_m<u^u \,,
\end{equation}
in order to ensure that the corresponding resolution of the jumps from $u^{u,d}$ back to the
undisturbed equilibrium $u=0$ is made by structures which propagate upstream and downstream respectively.
Equation (\ref{3-7}) supplemented by this condition determines the roots
$u^{u,d}$ without any ambiguity provided $\Delta$ satisfies the condition of their reality.
Indeed, typical graphs of the function of $F(u)$ for its characteristic parameters
are shown in Fig.~3  for the case $\al>0$. Similar graphs can be readily
plotted for the case $\al<0$.
\begin{figure}
\begin{center}
\includegraphics[width=6cm,clip]{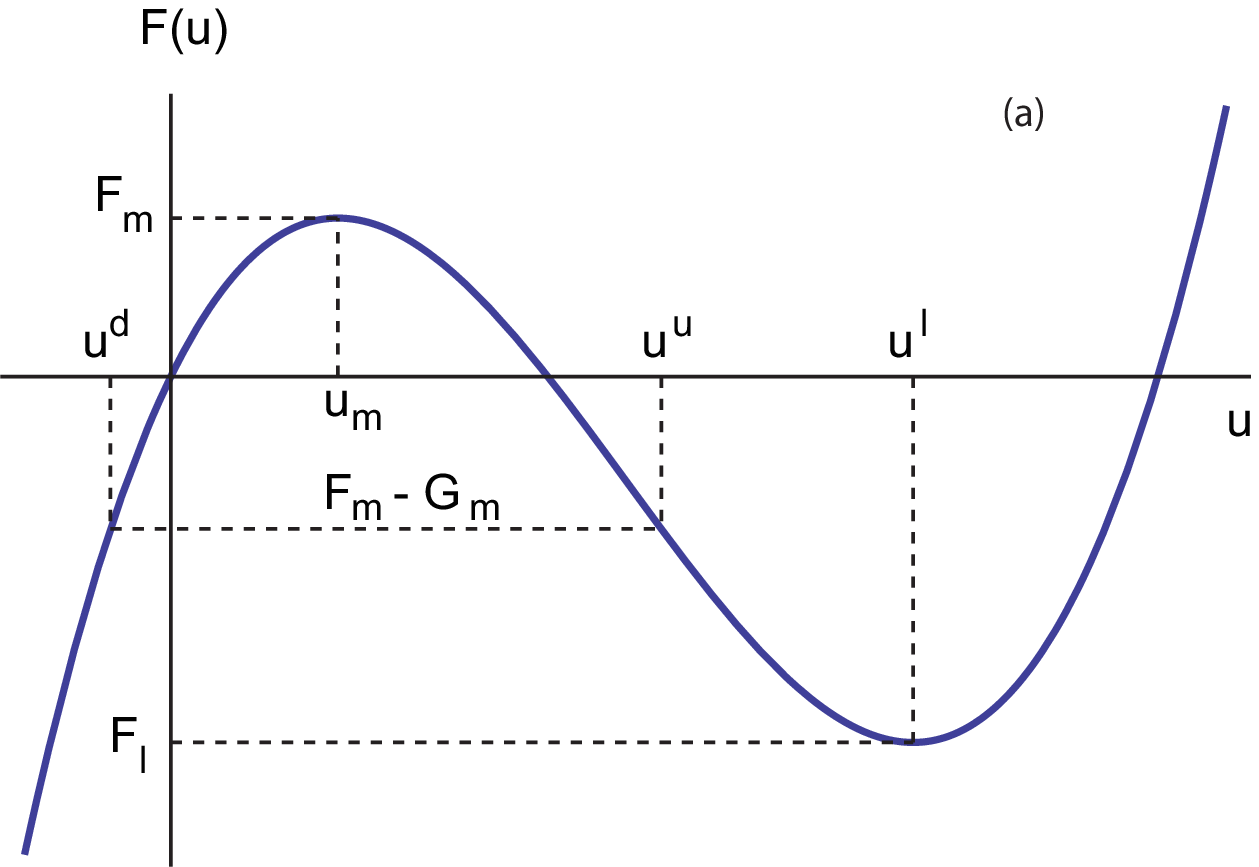} \qquad
\includegraphics[width=6cm,clip]{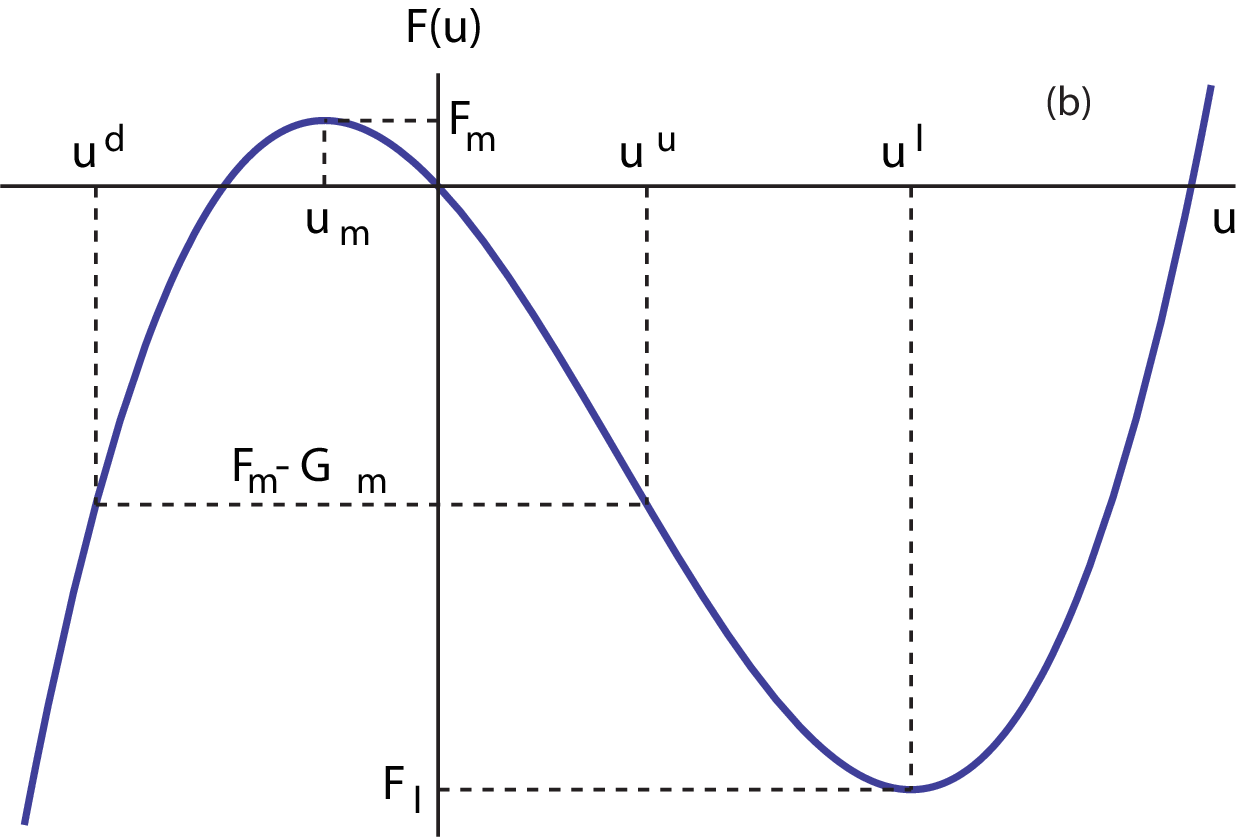}
\caption{Plots of the function $F(u)$ with $\al>0$ for (a) $\Delta>0$; and (b) $\Delta<0$.
 }
\end{center}\label{fig3}
\end{figure}

It is clear that equation (\ref{3-7}) has two real roots  $u^u$ and $u^d$ if $G_m$ satisfies the following condition
\begin{equation}\label{3-1a}
    F_m-G_m>F_l \,,
\end{equation}
where $F_m$ is the local maximal value (\ref{2-5}) and $F_l$ is the local minimal value (\ref{2-5b}),
which yields
\begin{equation}\label{3-3a}
    G_m<\frac1{\al^2}\left(1-\frac{2\al\Delta}3\right)^{3/2}.
\end{equation}

It is readily shown that the transcritical regime is indeed $\Delta_{-} < \Delta < \Delta_{+}$ where
the limiting values $\Delta_{\pm} (G_m )$ are precisely those found in section 2.1 as the lower and upper boundaries of
the supercritical and subcritical regimes respectively.
These boundaries are shown in the transcritical flow-regime diagrams in Figs.~4a and 4b by thick lines.
We note that the diagrams in Fig.~4 are universally presented in terms of $\al^2G_m$ and $|\al|\Delta$ so that they
do not depend on the stratification parameter $\alpha$.
We also mention that the ``cut-off'' of the upper  boundary of the transcritical region at some  $|\al|\Delta>0$
is consistent with the
behaviour of transcritical flow-regime diagram for the fully nonlinear transcritical two-layer flows
(see e.g. \cite{baines84} and \cite{wh2012}).

Next, from Eq.~(\ref{3-3a}) we conclude that the transcritical region is split into two subregions
by the line $\Delta'_+$ for $\al>0$ or by the line $\Delta'_-$ for $\al<0$ determined by the equations
\begin{equation}\label{3-3b}
    G_m=\frac1{\al^2}\left(1-\frac{2\al\Delta}3\right)^{3/2}\quad\text{or}\quad
    \Delta'_{\mathrm{sgn}(\al)}=\frac3{2\al}\left[1-(\al^2G_m)^{2/3}\right].
\end{equation}
These boundaries are shown in Fig.~4 by thick dashed lines. They appear for $\al^2G_m>1/8$ and mean that,
if $\al>0$ then in the domain $\Delta_-<\Delta<\Delta'_+$ both roots $u^{u,d}$ exist
whereas in $\Delta'_+<\Delta<\Delta_+$ the root $u^u$ disappears and the localised hydraulic solution
with different limiting values at $x\to\pm\infty$ ceases to exist. In a similar way, if $\al<0$, then
both roots $u^{u,d}$ exist for $\Delta'_-<\Delta<\Delta_+$, which corresponds to the existence of the
hydraulic solution with different asymptotic values at $x\to\pm\infty$, but such a solution does not exist
if  $\Delta_-<\Delta<\Delta'_-$.  This is in contrast with the forced KdV equation dynamics (see \cite{gs-1986}), where
the steady localised hydraulic solution exists within entire transcritical region (although the time of
establishment of this steady solution could be very long for the regimes close to the boundaries of the transcritical region).
\begin{figure}
\begin{center}
\includegraphics[width=6cm,clip]{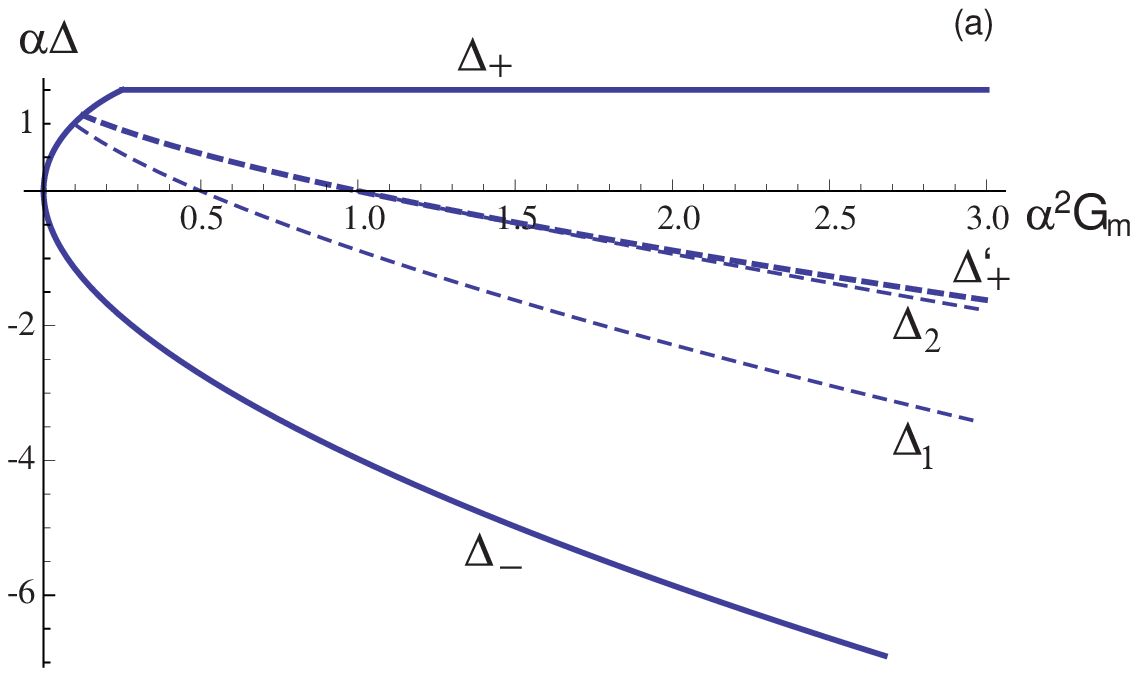} \qquad \quad
\includegraphics[width=6cm,clip]{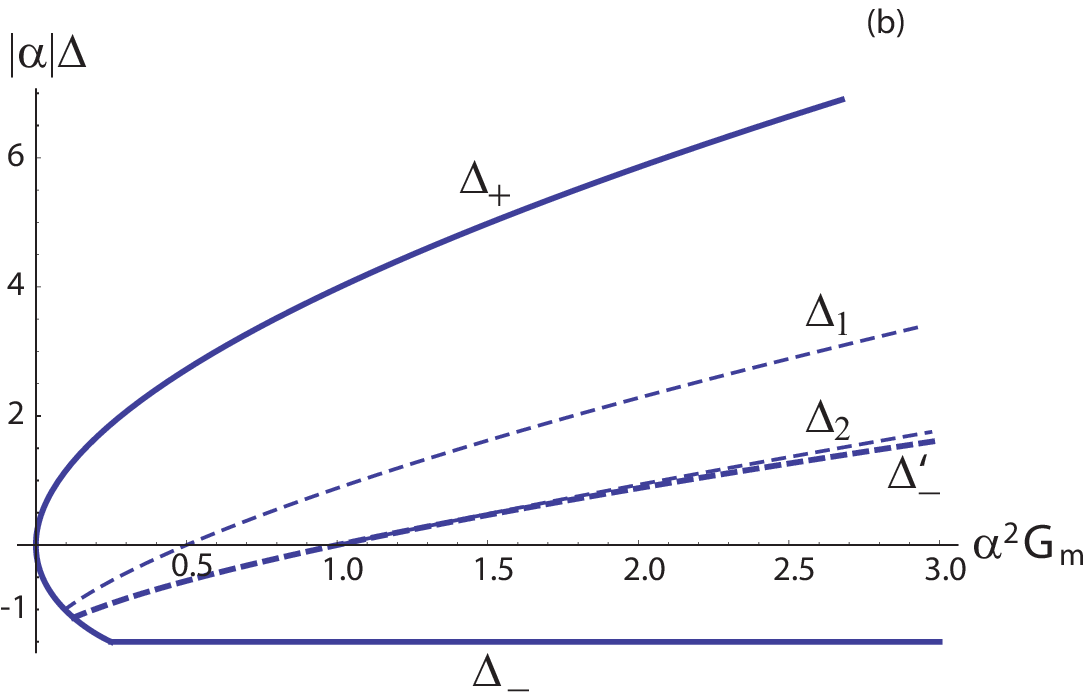}
\caption{Universal transcritical flow-regime diagrams in terms of $\al^2G_m$ and $|\al|\Delta$ for
 (a) $\al>0$, and (b) $\al<0$. }
\end{center}\label{fig4}
\end{figure}

Typical behaviours of $u^u,\,u^d$ as functions of $\Delta$ at a fixed $G_m$ are illustrated in Fig.~5
(note that the plots shown in Fig.~5 correspond to $G_m \alpha^2 < 1/8$ so the transcritical hydraulic
solution is valid in the whole range $\Delta_-<\Delta <\Delta_+$ -- see Fig.~4).
\begin{figure}
\begin{center}
\includegraphics[width=6cm,clip]{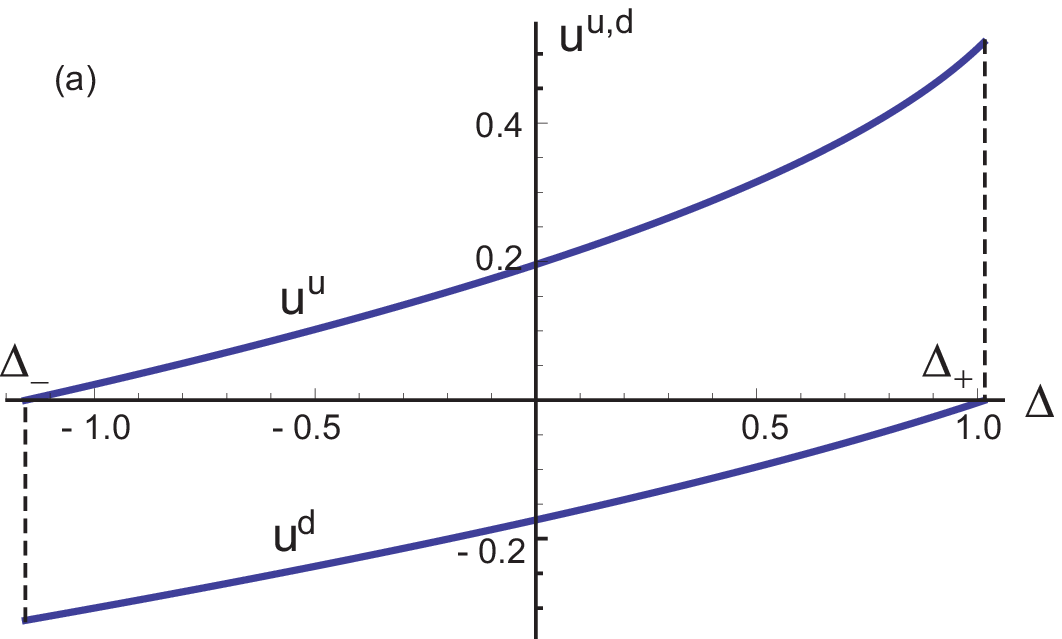} \qquad \quad
\includegraphics[width=6cm,clip]{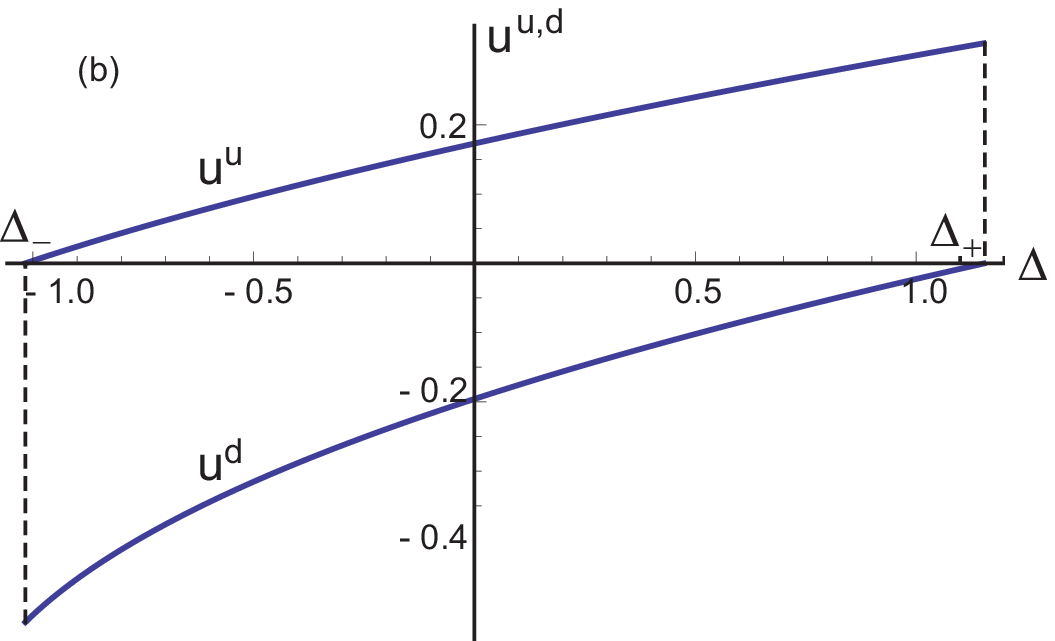}
\caption{Typical plots of $u^u$ and $u^d$ as functions of $\Delta$ for (a) $\al=+1$; and (b) $\al=-1$.
In both cases the forcing amplitude $G_m=0.1$.
 }
\end{center}\label{fig5}
\end{figure}
In the case of small $G_m$ we have the series expansions for $u^{u,d}$,
\begin{equation}\label{4-2a}
    u^{u,d}=\frac16\left[\Delta\pm\sqrt{12G_m}+\frac{\al}6\left(\Delta^2\pm\Delta\sqrt{12G_m}+4G_m\right)\right]+\ldots
\end{equation}
where the upper sign corresponds to $u^u$ and the lower one to $u^d$. The first two terms in (\ref{4-2a})
agree with the KdV approximation of \cite{gs-1986}.

When $u^u$ and $u^d$ as solutions of the equation (\ref{3-7}) are known, we are in position to construct
the unsteady solutions connecting these values with the equilibrium state $u=0$ at $x\to\pm\infty$ and
providing thus a closure for the localised transcritical hydraulic transition. Importantly, since the
corresponding unsteady structures are generated
away from the topography forcing,  their description is to be made in the framework of the homogeneous,
unforced Gardner equation. More specifically, we shall be interested in
the asymptotic solutions to the initial step problem for the Gardner equation assuming that for
sufficiently large times one can neglect the width of the obstacle compared with the width of the
generated expanding wave structure.

As is well known,  a single-layer shallow-water flow in the transcritical regime generates  upstream and downstream
undular bores, which in the weakly nonlinear approximation are asymptotically described by the modulated
periodic solutions of the (unforced) KdV equation; one of the bores being attached to the obstacle and
realised only partially (see \cite{gs-1986} and \cite{smyth-1987}).  In contrast to the KdV equation,
the Gardner equation supports a broad variety of unsteady solutions occurring in the step Riemann problem,
which, along with the KdV type ``cnoidal'' bores, include classical smooth bores (solibores), rarefaction waves,
nonlinear trigonometric bores and compound solutions representing combinations of the above structures.
These solutions were recently constructed by the present authors in \cite{kamch-12}, and below we shall
take advantage of the results of that work to describe weakly nonlinear unsteady wave regimes occurring
in transcritical  flows.

As follows from the results of \cite{kamch-12}, the type of the solution to the force-free Gardner equation
arising in the resolution of an initial step
\begin{equation}\label{3-4b}
    u(x,0)=\left\{
    \begin{array}{cc}
    {u}^- &  \hbox{for}\quad x<0,\\
    {u}^+ & \hbox{for} \quad x>0\,,
    \end{array}
    \right.
\end{equation}
strongly depends on the positions of boundary values $u^{\pm}$ relative to the turning point $u=1/(2\alpha)$
of the dispersionless limit characteristic velocity $c(u)=\Delta-6u(1-\alpha u)$ of the Gardner equation
(\ref{1-1}). We shall outline the Riemann problem classifications for $\alpha>0$ and $\alpha <0$ in the next section;
here we only take advantage of some basic results from \cite{kamch-12} enabling one to  identify the subdomains
within the transcritical regions $\Delta_-< \Delta <\Delta'_+$ $(\alpha>0)$ and $\Delta'_-< \Delta <\Delta_+$
$(\alpha<0)$ corresponding to qualitatively different behaviours of the unsteady solutions away from the obstacle.

If $\al>0$ then downstream of the obstacle one always has $u^-=u^d<0, u^+=0$, which corresponds to the cnoidal
undular bore of `normal' polarity (i.e. having a `bright' solitary wave of elevation at the leading edge)
propagating {\it to the right} (note that the Gardner equation (\ref{1-1}) with zero forcing term is related
to the Gardner equation in the standard form used in \cite{kamch-12} by a simple linear transformation of the
independent variables, see the next section).
However, upstream of the obstacle several different regimes are possible depending on the value of $u^u$.
These regimes are separated  by the lines in the $(\al^2G_m,\al\Delta)$-plane corresponding to $u^u=1/(2\al)$ and $u^u=1/\al$.
One can readily find that the line $u^u=1/(2\al)$ is determined by the equation
\begin{equation}\label{3-3c}
    G_m=\frac1{2\al^2}\left(1-\frac{2\al\Delta}3\right)^{3/2}\quad\text{for}\quad \al\Delta<1.
\end{equation}
This line is marked in Fig.~4a as $\Delta_1$. The line $u^u=1/\al$ is determined by the equation
\begin{equation}\label{3-3d}
    G_m=\frac1{2\al^2}\left[\left(1-\frac{2\al\Delta}3\right)^{3/2}+1\right] -\frac{\Delta}{2\al}
    \quad\text{for}\quad \al\Delta<0
\end{equation}
This line is marked in Fig.~4a as $\Delta_2$. Here the inequality $\al\Delta<0$ follows from the condition that,
for $u=u^u=1/\alpha$ to be a physical  root of the cubic equation (\ref{3-7}), the third (greatest) root of this
equation must be greater than $1/\al$. Note that (\ref{3-3d}) implies that for
$\alpha \Delta = 0$ one has $G_m \alpha^2=1$ so the line $\Delta_2$ starts at the upper boundary $\Delta_+'$ of
the existence of the hydraulic transcritical solution (see (\ref{3-3b})).

In a similar way, for $\al<0$ we see that for the upstream flow one has $u^-=0$ and $u^+=u^u>0$, which corresponds to the
`normal' undular bore with the soliton of positive polarity at the leading edge (see \cite{kamch-12} and section 3 below).
Downstream the obstacle three different regimes are possible depending on the value of $u^d$. The lines $u^d=1/(2\al)$
and $u^d=1/\al$ separating the regions for different regimes on the solution map in Fig.~4b are marked as
$\Delta_1$ and $\Delta_2$. Along the line $\Delta_2$ the third (smallest) root of the equation must (\ref{3-7})
lie outside the
segment $[u^d=1/\al; u^u]$, hence, we get $|\al|\Delta>0$ (cf. a similar restriction for $\alpha>0$ in (\ref{3-3d})).

We stress that the universal transcritical flow-regime classifications above are constructed in terms of
just two parameters $G_m \alpha^2$ and $|\alpha|\Delta$ while the original problem is specified by three parameters
$\alpha$, $G_m$ and $\Delta$. This means that the same upstream-downstream resolution
pattern can be realised for two disparate flows characterised by the same values of $G_m \alpha^2$ and $|\alpha|\Delta$.
Of course this property of transcritical stratified flows does not imply any non-uniqueness in the flow characterisation
for a given fixed value of the stratification parameter $\alpha$ but could be useful in the experimental modelling
where certain configurations are more accessible than others.

\section{Dispersive regularisation of  transcritical hydraulic jumps}

Before we proceed with the analytical description of the  wave patterns generated in
transcritical internal wave flows governed by the forced Gardner equation (\ref{1-1}) we review
the general classification of solutions to the initial step resolution (Riemann) problem described in our work,  \cite{kamch-12}.
This will enable us to identify the particular solutions realised in the dispersive regularisation of the upstream and
downstream hydraulic jumps forming in the transcritical flow.
An important feature of these solutions is that, despite the unidirectional nature of the Gardner equation the solution
of the step resolution problem generally consists of two parts (waves) which are pieced together. This is in sharp
contrast with the KdV equation dynamics where the resolution of an initial step always occurs
through a single wave (an undular bore or a rarefaction wave, see \cite{GP74}). The underlying mathematical reason
behind this broader variety of solutions is the fact that the modulation system for the Gardner equation,
unlike that for the KdV equation, is neither strictly hyperbolic (for $\alpha <0$) nor genuinely nonlinear
(for both signs of $\alpha$), which brings a suggestive parallel with shock theory in hyperbolic conservation
laws with non-convex flux (see e.g. \cite{leFloch}).
The key components of the asymptotic wave patterns occurring in the step resolution problem for the Gardner
equation are: undular bores (there are three different types of them), rarefaction waves and classical smooth bores (solibores).

\subsection{Undular bores: the ``core''  modulation solution}

We describe first the  ``core'' modulation solution for slow variations of the parameters of the periodic wave solution (such as amplitude, frequency, mean value) in the undular bore solutions to the unforced Gardner equation. This modulation solution plays the key role in the analytical description of the upstream and downstream wave patterns. As we have already mentioned,  the Gardner equation has several types of undular bore solutions realised for different sets of the initial step parameter values $u^-, u^+$. One of the main results of \cite{kamch-12} is that all modulation solutions for the Gardner equation `cnoidal' undular bores can be mapped onto the same classical Gurevich-Pitaevskii modulation solution for the KdV equation (see \cite{GP74}, \cite{fw78}). We would like to stress that this mapping is exact and is valid within the entire domain of the existence of the relevant undular bore solutions and thus, is different from an approximate near-identity transformation used in \cite{ms-1990} to describe the weakly nonlinear transcritical  regimes of forced Gardner equation when the higher nonlinear term is small and the upstream and downstream undular bore solutions are asymptotically close to those of the KdV equation.

The mapping between the Gardner and the KdV modulation solutions is, however, not one-to-one so the `physical' modulations in the Gardner undular bores are reconstructed using  one of the inverse mappings between the parameters of the relevant periodic solution of the Gardner equation  and the modulation variables in the Gurevich-Pitaevskii solution. The choice of the inverse mapping is determined by the sign of $\alpha$ and the position of the initial step parameters $u^-, u^+$ relative to the turning point $u=1/(2\alpha)$ of the dispersionless limit velocity $c(u)=\Delta-6u(1-\alpha u)$ of the Gardner equation (\ref{1-1}).

Another distinctive feature of the Gardner modulation dynamics is that for $\alpha<0$ it admits a special type of undular bores, termed nonlinear trigonometric bores, which have no analogue in the Gurevich-Pitaevskii theory for the KdV equation. These trigonometric undular bore solutions were first found in the focusing modified KdV equation (\cite{march2008}) and complex modified KdV equation (\cite{kod2008}).  A detailed description of trigonometric bores for the Gardner equation was made in \cite{kamch-12}.

All periodic solutions of the Gardner equation satisfy the ordinary differential equation (ODE)
\begin{equation}\label{eq3}
    u_{\xi}^2=\alpha(u-u_1)(u-u_2)(u-u_3)(u-u_4),
\end{equation}
where $\xi=x-Vt -\xi_0$, $V$ being the phase velocity and $\xi_0$ an arbitrary initial phase (see \cite{kamch-12} for details). In the modulationally stable case of our interest the roots of the polynomial in the right-hand side of (\ref{eq3}) are ordered
\begin{equation}\label{eq4}
    u_1\leq u_2\leq u_3\leq u_4
\end{equation}
and satisfy the condition
\begin{equation}\label{11-1}
    \sum_{i=1}^4u_i=\frac2{\al} \, ,
\end{equation}
so only three of them are independent. It is still convenient to keep all four $u_j$'s in the
subsequent formulae to preserve  symmetry of the expressions.
The phase velocity $V$ is expressed in terms of $u_j$'s as
\begin{equation}\label{el3}
  V=\Delta-\al(u_1u_2+u_1u_3+u_1u_4+u_2u_3+u_2u_4+u_3u_4).
\end{equation}
Now, if the parameters $u_j$ of the periodic solution are allowed to vary slowly with $x,t$, one arrives at a set of constraints on their variations, which have the form of quasilinear partial differential equations called the Whitham modulation equations (\cite{whitham1, whitham2}). The Whitham equations can be obtained by averaging conservations laws of the dispersive evolution equation over the periodic solution family.

It is clear that the actual form of the periodic solution of the Gardner equation specified by ODE (\ref{eq3}) will depend on the sign of $\alpha$ and on the interval  $(u_i, u_{i+1})$ within which the variable $u$ oscillates. As a result, for each family of the periodic solutions the Whitham averaging procedure would yield its `own' modulation system for $\{u_j(x,t), j=1, \dots, 4\}$ (we recall that only three of $u_j$'s are independent so the  modulation system will consist of three equations). However, as was shown in \cite{kamch-12}, for each set $\{u_j(x,t)\}$, there is a  mapping
$\{u_1, u_2, u_3, u_4\} \mapsto \{r_1,r_2,r_3\}$, where $r_3(x,t) \ge r_2(x,t) \ge r_1(x,t)$ are the Riemann invariants of the  well-known KdV-Whitham modulation system  (obtained by the Whitham averaging of the KdV reduction of the Gardner equation, when $\alpha=0$, see \cite{whitham2}). This (quadratic) mapping has the form
\begin{equation}\label{ru}
 r_1=\frac{\alpha}4(u_1+u_2)(u_3+u_4), \ \
 r_2=\frac{\alpha}4(u_1+u_3)(u_2+u_4), \ \
 r_3=\frac{\alpha}4(u_2+u_3)(u_1+u_4)
\end{equation}
for $\alpha>0$ and
\begin{equation}\label{ru-a}
 r_1=\frac{\alpha}4(u_1+u_4)(u_2+u_3), \ \
 r_2=\frac{\alpha}4(u_1+u_3)(u_2+u_4), \ \
 r_3=\frac{\alpha}4(u_1+u_2)(u_3+u_4)
\end{equation}
for $\alpha<0$.
\begin{figure}
\begin{center}
\includegraphics[width=6cm,clip]{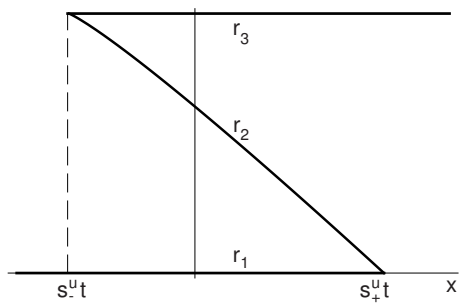}
\caption{Typical dependence of the Riemann invariants $r_i$ on the space coordinate $x$ in the Gurevich-Pitaevskii modulation solution (\ref{9-2}), (\ref{9-3}) for $r^-=0$, $r^+>0$ at some $t>0$.
 }
\end{center}\label{fig6}
\end{figure}

The similarity modulation solution of the KdV-Whitham system describing the dependence of $r_j$'s on $x$ and $t$
in the undular bore is well known (\cite{GP74}, \cite{fw78}) and has the form :
\begin{equation}\label{9-2}
r_1=r^-, \qquad r_3=r^+ \, ,
\end{equation}
where $r^+>r^-$ are constants, whereas the function $r_2(x,t)$, with an account of the coefficients
in the original Gardner equation (\ref{1-1}), is given implicitly by
\begin{equation}\label{9-3}
   \Delta-2(r_1+r_2+r_3)+\frac{4(r_2-r_1)(1-m)\K(m)}{\E(m)-(1-m)\K(m)}=\frac{x}{t},
\end{equation}
where $\K(m)$ and $\E(m)$ are the complete elliptic integrals of the fist and second kind respectively, and the modulus
\begin{equation}\label{8-3}
    m=\frac{r_2-r_1}{r_3-r_1}=\frac{(u_3-u_2)(u_4-u_1)}{(u_4-u_2)(u_3-u_1)}.
\end{equation}
The constants $r_1=r^-,r_3=r^+$ are determined by the initial step parameters $u^-,\,u^+$. These values also determine the specific set of relationships $u_j({\bf r})$ (one of the inverses to (\ref{ru}) or (\ref{ru-a})) to be used in the reconstruction of the Gardner undular bore solution from the dependence (\ref{9-3}) (see \cite{kamch-12}).

The modulated wavetrain described by (\ref{9-2}), (\ref{9-3}) is confined to an expanding region
\begin{equation}\label{int}
s^+t < x < s^-t \, ,
\end{equation}
where the leading and trailing edge speeds are
\begin{equation}\label{spm1}
s^+=\Delta  - 2 r^- - 4r^+\, , \qquad s^-=\Delta  - 12 r^- + 6r^+\,
\end{equation}
respectively. The values $s^{\pm}$ are found by substituting of $m=0$ (trailing edge) and $m=1$ (leading edge) in the similarity solution (\ref{9-3}).
The behaviour of the Riemann invariants $r_j(x,t)$ in the ``core'' modulation solution (\ref{9-2}), (\ref{9-3}) is shown in Fig.~6.
The inverse mappings $\{r_1, r_2,r_3\} \mapsto \{u_1,u_2,u_3,u_4\}$ for each relevant case will be described separately in sections 4 and 5.

\subsection{Riemann problem classification}

To describe the upstream and downstream closure of the localised transcritical transition described in section 2 we need to know solutions
of the Gardner equation (\ref{1-1}) with zero forcing term and steplike initial conditions (\ref{3-4b}). This (Riemann) problem  was considered in \cite{kamch-12}; here we present the resulting classifications of the solutions for $\alpha>0$ and $\alpha<0$. Note that
in \cite{kamch-12}  the Gardner equation was taken in the standard  form 
\begin{equation}\label{rG}
\tilde u_{\tilde t}+6\tilde u \tilde u_{\tilde x}-6\alpha \tilde u^2\tilde u_{\tilde x} +\tilde u_{\tilde x \tilde x \tilde x}=0\, ,
\end{equation}
which is related to the equation (\ref{1-1}) with $G_m \equiv 0$ by  a simple linear transformation $u=\tilde u, \quad t= \tilde t,  \quad x=-\tilde x + \Delta \tilde t$. The step conditions for (\ref{rG}) corresponding to (\ref{3-4b}) are
\begin{equation}\label{3-4a}
    \tilde u(\tilde x,0)=\left\{
    \begin{array}{cc}
    {u}^+, & \quad \tilde x<0,\\
    {u}^-, & \quad \tilde x>0\, .
    \end{array}
    \right.
\end{equation}

\begin{figure}
\centerline{\includegraphics[width=16cm, clip]{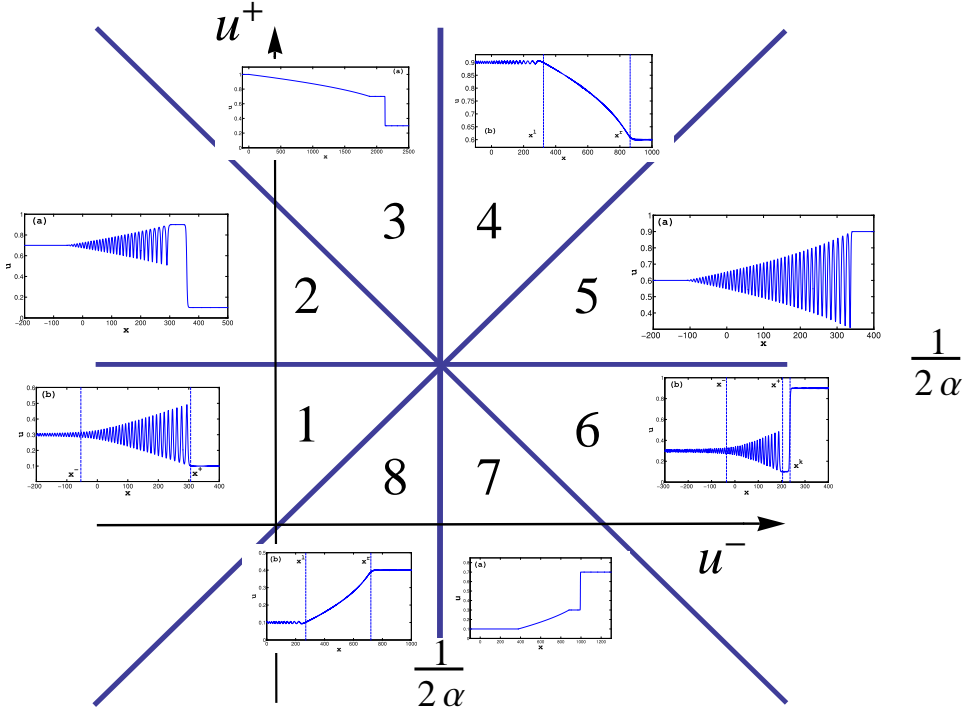}}
\caption{(Color online) Parametric map of solutions to the Gardner equation (\ref{rG}) with $\alpha>0$
and initial step conditions (\ref{3-4a}) (from \cite{kamch-12}).
The resolution diagrams corresponding to each of the cases on the plane of the initial step parameters
$u^+$ and $ u^-$ are the following:
Region 1: $\{ u^+  \ {\bf UB} \rightarrow  u^-\}$;
Region 2: $\{ u^+  \leftarrow {\bf UB} \ (\tilde u^*) \ {\bf SB} \rightarrow  u^-\}$;
Region 3: $\{u^+ \leftarrow {\bf RW} \ ( \tilde u^*) \ {\bf SB} \rightarrow  u^-\}$;
Region 4: $\{ u^+$ {\bf $\leftarrow$ RW } $ u^- \}$;
Region 5: $\{ u^+$ {\bf $\leftarrow$ UB } $ u^- \}$;
Region 6: $\{ u^+$ {\bf UB} $\rightarrow \ (\tilde u^*)$ $\leftarrow$ {\bf SB} $ u^-\}$;
Region 7: $\{u^+$ {\bf RW} $\rightarrow \ ( \tilde u^*)$ $\leftarrow$ {\bf SB} $ u^-\}$;
Region 8:  $\{ u^+$ {\bf RW $\rightarrow$ } $ u^-\}$.
In all relevant cases the intermediate state $\tilde u= \tilde u^*=1/\alpha -  u^-$.}
\label{classpos}
\end{figure}
\begin{figure}
\centerline{\includegraphics[width=16cm, clip]{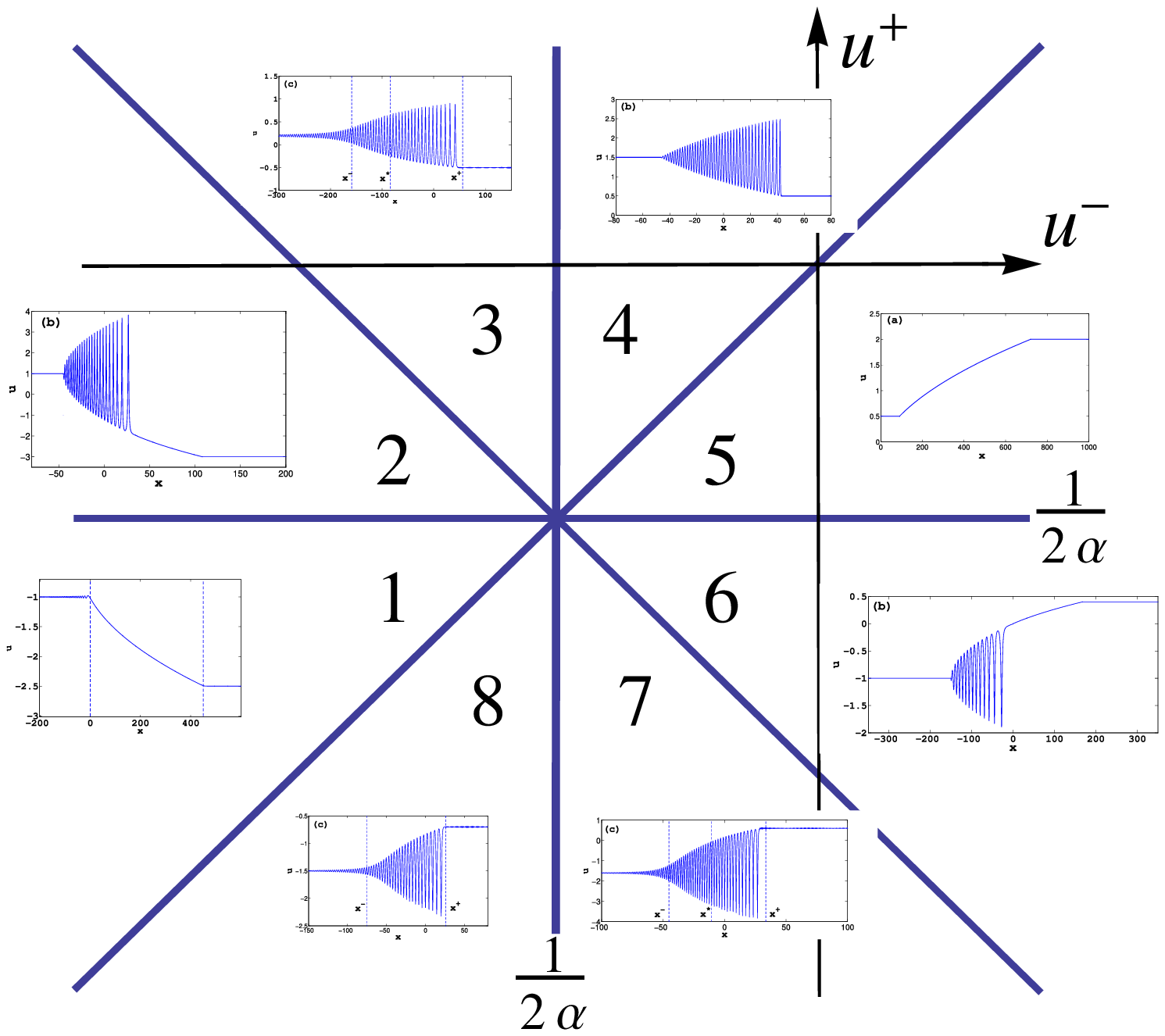}}
\caption{(Color online) Parametric map of solutions of the step problem for the Gardner equation (\ref{rG}) with
$\alpha<0$ and initial step conditions (\ref{3-4a}) (from \cite{kamch-12}). The resolution diagrams corresponding
to each of the cases on the plane of the initial step parameters
$u^-$ and $\tilde u^+$ are the following:
Region 1: $\{ u^+  \leftarrow {\bf RW} \  u^-\}$;
Region 2: $\{ u^+ \ {\bf TB} \rightarrow \ ( \tilde u^*)  \leftarrow  {\bf RW} \  u^-\}$;
Region 3:  $\{\tilde u^- \ ({\bf TB}|{\bf UB}) \rightarrow  \ \tilde u^+\}$;
Region 4: $\{u^+$ {\bf  UB } $\rightarrow$ $ u^- \}$;
Region 5: $\{u^+$ {\bf  RW } $\rightarrow$ $ u^- \}$;
Region 6: $\{u^+ \ \leftarrow{\bf TB}  \ (\tilde u^*)  \ {\bf RW} \rightarrow \  u^-\}$;
Region 7: $\{ u^+ \leftarrow  \ ({\bf TB} | {\bf UB}) \  u^- \}$;
Region 8:  $\{ u^+ \leftarrow {\bf UB} \  u^-\}$.
In all relevant cases the intermediate state $ \tilde u = \tilde u^*=1/\alpha -  u^+$. }
\label{classneg}
\end{figure}

\medskip
The classifications of solutions to the Riemann problem  (\ref{rG}), (\ref{3-4a}) with $\alpha>0$ and $\alpha<0$ are presented in Fig.~\ref{classpos} and Fig.~\ref{classneg}  respectively. Since in the transcritical flow  we always have $u^-=0$ for the upstream solution and $u^+=0$ for the downstream solution, only certain parts of the full classification are relevant to the problem under study, these will be detailed in sections 4 and 5. Here we only explain the notations for the step resolution diagrams used in the figure captions. The labels {\bf UB}, {\bf RW} and {\bf SB} denote the cnoidal undular bore, the rarefaction wave and the solibore respectively; the arrows denote the `normal' ($\rightarrow$) and `reversed' ($\leftarrow$)  structure of the relevant wave pattern. The terms `normal' and `reversed' refer to the sign of the gradient of $\tilde u$  across the structure: in the `normal' undular bores and solibores the value of $\tilde u$ decreases when crossing the structure in the positive direction, while in the `normal' rarefaction wave this value increases. In the `reversed' structures the signs of the gradients are opposite to those in `normal' ones.  A special feature of the reversed undular bores is that, despite the negative sign of linear dispersion in the Gardner equation, they exhibit {\it ``dark'' solitons} at their leading edges in contrast to the `normal' KdV-type negative dispersion ``bright soliton'' pattern.

For instance, the diagram $\{u^+ \leftarrow {\bf RW} \ (\tilde u^*) \ {\bf SB} \rightarrow  u^-\}$ in Fig.~\ref{classpos} (region 3) corresponds to the  resolution pattern for the initial step (\ref{3-4a}) consisting of  the `normal' solibore and the `reversed' rarefaction wave connected via an intermediate state $\tilde u^*=1/\alpha -  u^-$.  The nonlinear trigonometric bores appearing in the classification for $\alpha<0$ are denoted as {\bf TB}. In a number of cases, the trigonometric bore occurs as part of a composite structure: say, the resolution diagram $\{u^+ \leftarrow  \ ({\bf TB} | {\bf UB}) \  u^- \}$ in Fig.~\ref{classneg} (Region 7) denotes the composite cnoidal-trigonometric undular bore connecting the constant boundary states $u^-$ and $u^+$.

\bigskip
\section{Upstream and downstream closure, $\alpha>0$}

\subsection{Upstream solution}
To describe the upstream wave pattern we need to solve the unforced Gardner equation (\ref{1-1}) with
initial conditions (\ref{3-4b}) where
$u^-=0$ and $u^+=u^u$. The dependence of the upstream hydraulic elevation $u^u$ on the  parameters
$\Delta$, $G_m$ is given by (\ref{3-7}), (\ref{4-1}).
The value of $u^u$ within the region of existence of the localised transcritical hydraulic solution
(see Fig.~5a) ranges
from $(u^u)_{min}=0$ at the lower boundary $\Delta_-$  to some  $(u^u)_{\max}$ at the upper boundary $\Delta'_+$.
The value $(u^u)_{\max}$ is found as the smaller positive root of equation (\ref{3-7}) with
$G_m(\Delta)$ defined by (\ref{3-3b}).

An inspection of the classification presented in Fig.~\ref{classpos} shows that within this range of $u^+=u^u$
(at fixed $u^-=0$) one can possibly have three different types of the solution corresponding to Regions 1,2 and 3.
These solutions are presented below.

 \medskip

(a) $\Delta_- <\Delta <\Delta_1$ (see Fig.~4a).  This corresponds to $0<u^u<1/(2\al)$ and, hence, to Region 1 in  the classification map in Fig.~\ref{classpos}. Thus, we have an upstream propagating undular bore with the bright soliton at the leading (left) edge (note the change in the propagation direction when applying the classification in Fig.~\ref{classpos} to equation (\ref{1-1})). The modulation solution for this bore is given by (\ref{9-2}), (\ref{9-3}), where the constant Riemann invariants (\ref{9-2}) are
\begin{equation}\label{9-4}
    r_1=0,\quad r_3=u^u(1-\al u^u),
\end{equation}
and the dependence of the varying Riemann invariant $r_2$ on $x$ and $t$ is given in an implicit form by (\ref{9-3}).

The oscillatory structure of the undular bore is described by the periodic solution of the ODE (\ref{eq3}), which is  expressed in terms of the Jacobi elliptic functions as (see \cite{kamch-12})
\begin{equation}\label{5-3}
    u=u_2+\frac{(u_3-u_2)\mathrm{cn}^2(\theta,m)}{1-
    \frac{u_3-u_2}{u_4-u_2}\mathrm{sn}^2(\theta,m)} \, ,
\end{equation}
where the phase is
\begin{equation}\label{5-4}
    \theta=\sqrt{\al(u_3-u_1)(u_4-u_2)}(x-x_0-Vt)/2 \, ,
\end{equation}
the modulus $m$ is given by (\ref{8-3}),
and the phase velocity $V$ is
\begin{equation}\label{5-6}
    V=\Delta-\al(u_1u_2+u_1u_3+u_1u_4+u_2u_3+u_2u_4+u_3u_4).
\end{equation}
In the small-amplitude limit $u_3\to u_2$ ($m_1\to 0$) we get the linear harmonic wave
\begin{equation}\label{5-11}
\begin{split}
    &u\cong u_2+\tfrac12(u_3-u_2)\cos[k(x-x_0-Vt)],\\
    &k=\sqrt{\alpha(u_2-u_1)(u_4-u_2)},\\
    &V=\Delta-4u_2-\alpha(u_1u_4-3u_2^2).
    \end{split}
\end{equation}
The soliton limit $m \to 1$ can be achieved in one of the two ways: when $u_1 \to u_2$ or when $u_3 \to u_4$.
In the Region 1 the soliton limit occurs via $u_2 \to u_1$ (see \cite{kamch-12}) so we obtain a ``bright''
soliton of elevation propagating against a constant background $u=u_1$,
\begin{equation}\label{5-8}
    u(\theta)=u_1+\frac{u_3-u_1}{\cosh^2\theta-
    \frac{u_3-u_1}{u_4-u_1}\sinh^2\theta} \, .
\end{equation}
If, further, one has $u_4-u_3 \ll u_3-u_2,$ then the soliton (\ref{5-8}) becomes a wide,
``table-top" soliton.

For Region 1, the expressions for the parameters $u_j$ in terms of the Riemann invariants $r_1, r_2, r_3$
are given by the following inverse formulae of (\ref{ru-a}) (see \cite{kamch-12}),
\begin{equation}\label{8-4}
    \begin{split}
    &u_1=\frac1{2\al}\left(1-\sqrt{1-4\al r_1}-\sqrt{1-4\al r_2}+\sqrt{1-4\al r_3}\right),\\
    &u_2=\frac1{2\al}\left(1-\sqrt{1-4\al r_1}+\sqrt{1-4\al r_2}-\sqrt{1-4\al r_3}\right),\\
    &u_3=\frac1{2\al}\left(1+\sqrt{1-4\al r_1}-\sqrt{1-4\al r_2}-\sqrt{1-4\al r_3}\right),\\
    &u_4=\frac1{2\al}\left(1+\sqrt{1-4\al r_1}+\sqrt{1-4\al r_2}+\sqrt{1-4\al r_3}\right),
    \end{split}
\end{equation}
which, after the substitution of (\ref{9-4}), assume the form (note that $(1-4\alpha r_3)^{1/2} = 1-2\alpha u^u$
for $0<u^u < 1/(2\alpha)$ -- see (\ref{9-4})):
\begin{equation}\label{9-4a}
    \begin{split}
    u_1(r_2)&=\frac1{2\al}(1-2\al u^u-\sqrt{1-4\al r_2}),\\
    u_2(r_2)&=\frac1{2\al}(-1+2\al u^u+\sqrt{1-4\al r_2}),\\
    u_3(r_2)&=\frac1{2\al}(1+2\al u^u-\sqrt{1-4\al r_2}),\\
    u_4(r_2)&=\frac1{2\al}(3-2\al u^u+\sqrt{1-4\al r_2}).
    \end{split}
\end{equation}
One can see that the limit $m=1$ (i.e. $r_2=r_3=u^u(1-\al u^u)$) is indeed achieved via  $u_2=u_1=0$ (see (\ref{8-3})),
which corresponds to the ``bright''  soliton (\ref{5-8}). In the linear limit $m=0$ ($r_2=r_1=0$)  one has $u_3=u_4=u^u$ .
The qualitative behaviour of $u_j$'s (\ref{9-4a}) generated by the ``core'' solution (\ref{9-3}) in the normal undular
bore (Region 1 in Fig.~7) is shown in Fig.~9a.

The speeds of the leading (soliton) and the trailing (linear) edges of the upstream undular bore edges are given by (see (\ref{spm1}), (\ref{9-4}))
\begin{equation}\label{9-5}
    s^-_u=\Delta-4u^u(1-\al u^u),\qquad s^+_u=\Delta+6u^u(1-\al u^u)
\end{equation}
respectively. Formulae (\ref{9-5}), via $u^u({\Delta, G_m})$ specified by (\ref{3-7}), (\ref{4-1}) define
the undular bore edge speeds in terms of the input parameters $\Delta, G_m$. The amplitude of the lead soliton is $a^-=2u^u$.
\begin{figure}
\begin{center}
\includegraphics[width=6cm,clip]{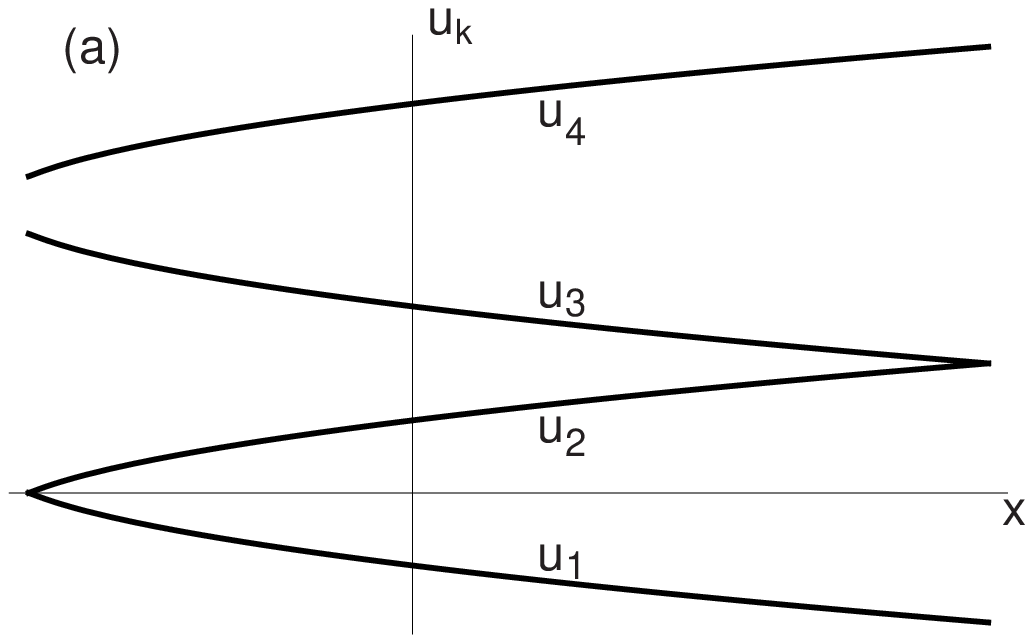} \qquad
\includegraphics[width=6cm,clip]{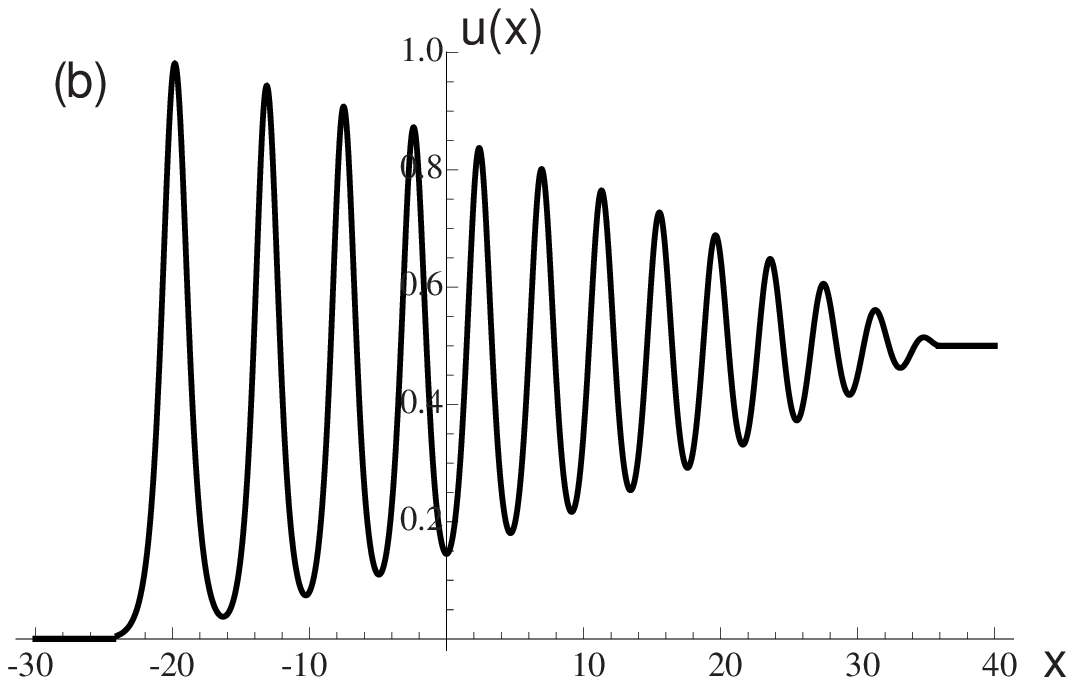}
\caption{Riemann problem solution to the  unforced Gardner equation (\ref{1-1}) with $\alpha >0$
and the initial step parameters  $u^-=0$, $u^+=u^u$, $0<u^u<1/(2\al)$,  corresponding to the upstream resolution of the transcritical flow via a normal undular bore
 (Region 1 in the Riemann problem classification in Fig.~7 and the region $\Delta_-< \Delta <\Delta_1$ of the transcritical flow-regime diagram in Fig.~4a).
(a)  Typical behaviour of the modulation parameters $u_k$  in the undular bore.
At the soliton edge $u_1=u_2=0$; at the linear edge $u_2=u_3=u^u$;\
(b) Plot of the analytical (modulation theory) solution to the  Gardner equation with $\al=1$, $\Delta=0$ and the step-like initial condition (\ref{3-4b}) with $u^-=0,$ $u^+=0.45$ at the time $t=40$.
Only the upstream part, $x<0$, of the bore is realised in the transcritical flow.
 }
\end{center}\label{fig9}
\end{figure}

\begin{figure}
\begin{center}
\includegraphics[width=8cm, clip]{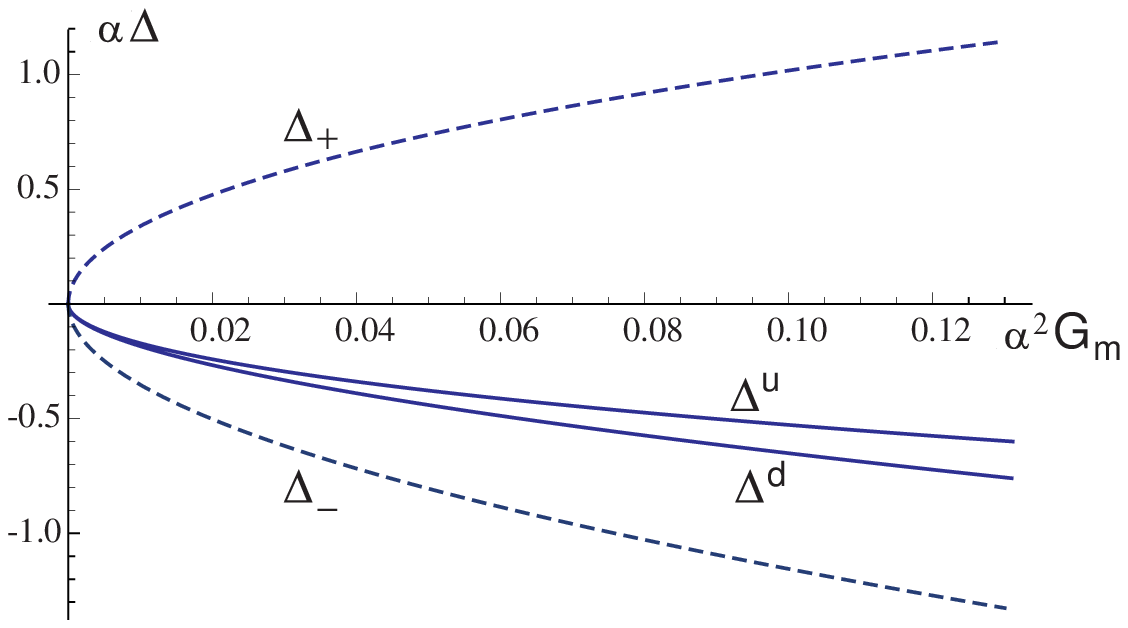}
\caption{Detailed structure of the transcritical flow-regime diagram for $\alpha \Delta <1$.
The lines $\Delta^u$ and $\Delta^d$ in the $(G_m \alpha^2, \Delta \alpha)$--plane separate
the sub-regions of the parameters corresponding to the generation of the attached (partial) and detached (fully realised) bores. The line
$\Delta^u$ is specified by (\ref{und-4}), the line $\Delta^d$ is specified by (\ref{und-5}).
The boundaries $\Delta_{\pm}$ of the transcritical region are shown by dashed lines. Specifically:
(i) $\Delta_-<\Delta<\Delta^d$ -- the upstream bore is detached, the downstream bore is attached;
(ii)  $\Delta^d<\Delta<\Delta^u$ -- both bores are detached;
(iii) $\Delta^u<\Delta<\Delta_+$ -- the upstream bore is attached, the downstream bore is detached.
For $\alpha \Delta >1$ the upper transcritical boundary $\Delta_+$ in (iii) should be replaced by $\Delta_1$ -- see Fig. 4a.
}
\end{center}
\label{fig10}
\end{figure}

It is clear that the upstream undular bore can be realised fully only if its trailing edge propagates upstream,
i.e. if  $s^+<0$. The condition $s^+=0$ yields then the equation of the line in the $( \alpha \Delta, G_m \alpha^2)$-plane
of the flow-regime diagram (see Fig.~4a) corresponding to the generation of an upstream undular bore whose
zero-amplitude trailing edge is at rest at the obstacle's location (effectively at $x=0$). Since $u=u^u$ is a solution
of (\ref{3-7}), an analytical expression for this line $\Delta=\Delta^u(G_m)$ is obtained by finding $u^u$ from
the equation $s^+=0$ specified by (\ref{9-5})  and then substituting it into (\ref{3-7}):
\begin{equation}\label{und-4}
   \Delta^u: \quad  G_m \alpha^2=\frac1{2}\left\{\frac12\left(1-\sqrt{1+\frac{2\al\Delta}3}\right)
    \left(1-\sqrt{1+\frac{2\al\Delta}3}-\frac{8\al\Delta}3\right)-1+\al\Delta+\left(1-\frac{2\al\Delta}3
    \right)^{3/2}\right\}.
\end{equation}
Expanding (\ref{und-4}) for small $|\alpha \Delta| \ll 1$ we obtain:
\begin{equation}\label{u-6}
 \Delta^u: \quad   G_m \alpha^2 \cong \frac{1}{3}(\alpha \Delta)^2 - \frac{1}{27}(\alpha \Delta )^3 \, .
\end{equation}
The inverse function  $\Delta(G_m)$ to (\ref{und-4}) has two branches; we are interested in the lower one
located within the region $\Delta_-<\Delta<\Delta_1$. Above this line one has $s^+>0$, and, therefore,
the upstream undular bore can be realised only partially in this region of the input parameters (see Fig.~10).
We also note that the leading term of the expansion (\ref{u-6}) $\Delta = - \sqrt{3 G_m}$ agrees with
the forced KdV result by \cite{gs-1986} and \cite{smyth-1987}, as expected. We also note that the separation
curve (\ref{und-4}) is defined only for  $\alpha \Delta>-3/2$ (the existence of the real root $u^u$ of
the quadratic equation $s^+=0$). For $\alpha \Delta<-3/2$ the upstream bore is always detached.

An example of the Riemann problem solution (\ref{5-3}), (\ref{9-4a}), (\ref{9-2}), (\ref{9-3}) corresponding
to the upstream flow resolution via partial bore is shown in Fig.~9b. Indeed, in the context
of the transcritical flow, the downstream part of the bore cannot be realised and one  ends up with the
partial undular bore attached to the obstacle at $x=0$ (as long as the  modulation description is concerned).
The partial undular bore has some nonzero value of the modulus $m=m_0>0$ (and, therefore, some non-zero
amplitude $a=a_0>0$) at the attachment point.  The  parameters $m_0$ and $a_0$ are found from the similarity
solution (\ref{9-3}) with $r_1=0$ complemented by the condition of continuity for the mean value $\bar u$ at $x=0$,
where one requires $\bar u = u^u$. The analysis is equivalent to that performed in \cite{gs-1986} and
\cite{smyth-1987} for the forced KdV equation so we do not present it here. We just mention that,
if $m_0$ is sufficiently close to unity one can replace the partial undular bore by a uniform solitary wave train.

\medskip
(b) $\Delta_1 <\Delta <\Delta_2$ (see Fig.~4a).  This corresponds to $1/(2\al)<u^u<1/\al$ and, hence,
to the Region 2 in  the parametric map of the Riemann problem in Fig.~\ref{classpos}. The upstream wave
structure now consists of the reversed undular bore with the dark soliton at the leading edge and a solibore
(kink) propagating to the left into the undisturbed state $u^-=0$ (see Fig.~11). The undular bore and the
solibore are joined together via an intermediate state $u=u^*=1/\alpha - u^-$.
\begin{figure}
\begin{center}
\includegraphics[width=6cm,clip]{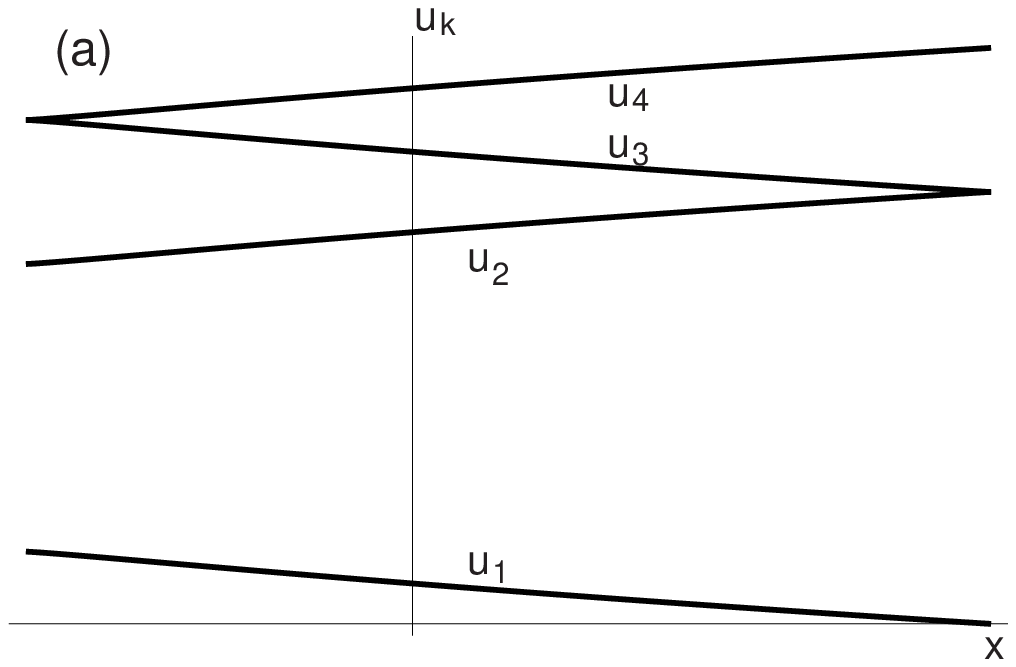} \qquad
\includegraphics[width=6cm,clip]{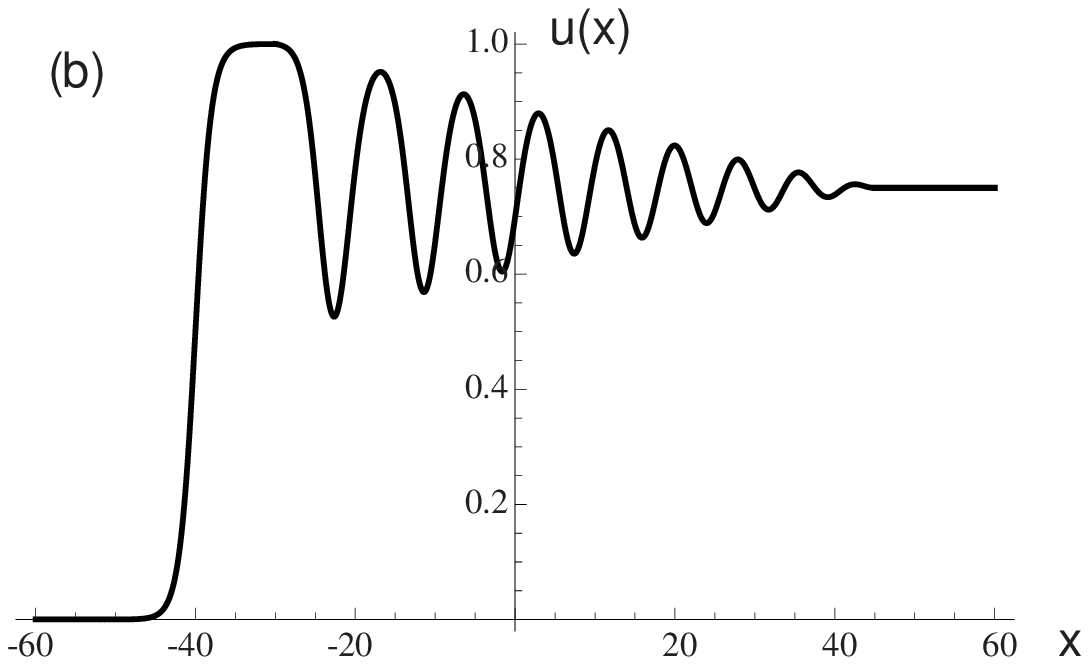}
\caption{Riemann problem solution to the  unforced Gardner equation (\ref{1-1}) with $\alpha >0$
and the initial step parameters  $u^-=0$, $u^+=u^u$, $1/(2\al)<u^u<1/\al$,  corresponding to the region $\Delta_1< \Delta <\Delta_2$ in the transcritical flow diagram in Fig.~4a (Region 2 in Riemann problem classification in Fig~7).
(a) Typical behaviour  of the modulation parameters $u_k$ in the reversed  undular bore.
At the leading (dark soliton) edge of the  bore $u_3=u_4=u^*=1/\alpha$; at the trailing (linear) edge $u_3=u_2=u^u<1/\al$.
(b) Plot of the analytical (modulation theory) solution to the  Gardner equation with $\al=1$, $\Delta=0$ and the
 step-like initial condition (\ref{3-4b}) with $u^-=0,$ $u^+=0.75$ at the time $t=40$. The constant state $u=u^*=1$ at the leading edge of the reversed undular bore is further connected with the equilibrium flow upstream  $u=0$ via a solibore (kink).}
\end{center}\label{umod}
\end{figure}

The reversed undular bore is described by the same ``core'' modulation solution (\ref{9-4}), (\ref{9-3})
in terms of the Riemann invariants $r_j$, $j=1,2,3$ but
the relationships between the parameters $\{u_j\}$ in the periodic solution (\ref{5-3}) and the Riemann invariants
$\{r_k\}$ are now given by the second set of the inverse formulae to (\ref{ru}),
\begin{equation}\label{8-5}
    \begin{split}
    &u_1=\frac1{2\al}\left(1-\sqrt{1-4\al r_1}-\sqrt{1-4\al r_2}-\sqrt{1-4\al r_3}\right),\\
    &u_2=\frac1{2\al}\left(1-\sqrt{1-4\al r_1}+\sqrt{1-4\al r_2}+\sqrt{1-4\al r_3}\right),\\
    &u_3=\frac1{2\al}\left(1+\sqrt{1-4\al r_1}-\sqrt{1-4\al r_2}+\sqrt{1-4\al r_3}\right),\\
    &u_4=\frac1{2\al}\left(1+\sqrt{1-4\al r_1}+\sqrt{1-4\al r_2}-\sqrt{1-4\al r_3}\right).
    \end{split}
\end{equation}
Upon substituting (\ref{9-4}) and using that  $(1-4\alpha r_3)^{1/2} = 2\alpha u^+-1$ (now $u^+>1/(2\alpha)$),
one can readily see that the relationships (\ref{8-5}) assume the same form (\ref{9-4a}). However, due to the
different sign of $(1-4\alpha r_3)^{1/2}$, the soliton limit $r_2=r_3$ ($m=1$) is now achieved via $u_3=u_4$,
which corresponds to the {\it dark} soliton limit of the periodic solution (\ref{5-3}) (see \cite{kamch-12} for details):
\begin{equation}\label{5-9}
    u(\theta)=u_4-\frac{u_4-u_2}{\cosh^2\theta-
    \frac{u_4-u_2}{u_4-u_1}\sinh^2\theta}.
\end{equation}
Now, it follows from (\ref{8-5}) that, when $r_2$ varies in the interval $r_1\leq r_2\leq r_3$ according to (\ref{9-3}),
the parameters $u_k$ change along the curves shown in Fig.~11a (cf. Fig.~9a for the Region 1 undular bore).
The left degeneration point $u_3=u_4=1/\alpha$ corresponds to a dark soliton (\ref{5-9}) riding on the background
$u=u^*=1/\al>0$, while at the opposite  edge  we have $u_3=u_2=u^+=u^u$, which corresponds to a linear wave packet
propagating on the uniform background $u=u^u$. Thus, substitution of $u_k(x,t)$ in the periodic solution (\ref{5-3})
yields the reversed undular bore connecting the levels $u=u^u$ and $u=1/\alpha$ (see Fig.~11b). The lead soliton
amplitude in this undular bore is $a^*=2(1/\alpha - u^u)$.

The intermediate state $u=1/\al$ at the leading edge is further connected with the upstream equilibrium state $u=0$
by a solibore, which is obtained as the limiting form of the periodic solution (\ref{5-3}) when
$u_2 \to u_1$ and $u_3 \to u_4$ simultaneously. There are two possible solibore solutions, the one relevant to
Region 2 has the form
\begin{equation}\label{5-10}
u(x,t)=\frac{1}{\alpha}\left(1-\frac{1}{\exp[\sqrt{\alpha}(x-s^k t)]+1}  \right)  \, , \qquad  s^k=\Delta-1/\al.
\end{equation}
The  type of the initial step resolution via the combination of an undular bore and a solibore described above
is sometimes referred to as the dispersive
``lock-exchange'' (in contrast to the dispersive ``dam-break''  type resolution in Region 1, see for instance \cite{ep-11}.
The speeds of the soliton and linear edges of the reversed upstream undular bore in the lock-exchange solution are  given
by the same formulae (\ref{9-5}) which, for the region $\Delta_1 <\Delta <\Delta_2$ considered, implies that the reversed
bore is  attached to the obstacle and realised only partially  as long as $ \alpha \Delta > -3/2$ (see (\ref{und-4})).
We note that, although this pattern can be seen in the previous numerical simulations (see e.g. \cite{gcc-2002})
it had not been identified as a separate distinct feature of the dispersive transcritical stratified flows.

\medskip

(c) $\Delta_2 <\Delta <\Delta'_+$ (see Fig.~4a; note that this narrow region is  located entirely in the lower half-plane, i.e. for $\alpha \Delta  <0$, $G_m \alpha^2>1$).  This corresponds to $1/\al<u^u<u^u(\Delta'_+)$ and, hence, to the Region 3 in  the parametric classification of the Riemann problem in Fig.~7. The upstream wave structure now consists of the combination of the solibore and the reversed rarefaction wave  connected via an intermediate state $u=u^*=1/\al$.
This is the classical lock-exchange pattern observed in previous numerical simulations of the forced Gardner equation
(see e.g. \cite{mh-1987}, \cite{gcc-2002}) and also in fully nonlinear two-layer transcritical flows \cite{wh2012}.
\begin{figure}
\begin{center}
\includegraphics[width=7cm, clip]{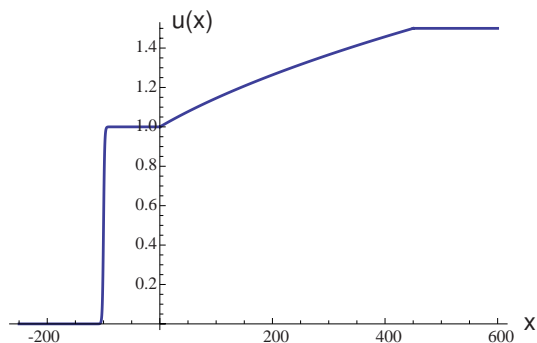}
\caption{Asymptotic solution (\ref{5-10}), (\ref{revrare}) to the Riemann problem for the unforced Gardner equation (\ref{1-1}) with $\alpha>0$ and the initial step parameters $u^-=0$, $u^+=u^u$, where $1/\al<u^u<u^u(\Delta'_+)$ (Region 3 in the Riemann problem classification Fig~7 and the corresponding region $\Delta_2< \Delta <\Delta_+'$ in the transcritical flow-regime diagram in Fig.~4a).   The pattern consists of the rarefaction wave (exact solution of the dispersionless Gardner equation)  and the solibore joined together via the constant state $u=1/\alpha$. The plot parameters are: $\alpha =1$, $\Delta=0$, $u^-=0$, $u^+=u^u=1.5$, $t=100$.}
\end{center}\label{fig12}
\end{figure}
The solibore connecting the equilibrium state $u=0$ upstream and the intermediate constant state $u=1/\alpha$ is described by formula (\ref{5-10}). The constant state is further connected with the upstream hydraulic elevation state $u=u^u$ via the reversed simple rarefaction wave, which is asymptotically described by the similarity solution of the dispersionless limit of the unforced Gardner equation,
\begin{eqnarray}\label{revrare}
u &=& 1/\alpha,  \quad \hbox{for}  \quad   x < s^l  t \, ,\nonumber \\
u &=& \frac1{2\al}\left(1+\sqrt{1+\frac{ 2\alpha } {3}
\left(\frac{x}{t} - \Delta\right)}\right)  \quad \hbox{for}
\quad  s^l t < x <  s^r t \,, \nonumber \\
 u &=& u^u \quad \hbox{for} \quad x>s^r t   \, ,
\end{eqnarray}
where the edge speeds are
\begin{equation}\label{rrw}
s^l=\Delta\, , \qquad s^r=\Delta - 6u^u(1-\alpha u^u)\, .
\end{equation}
As a matter of fact, $s^k<s^l$ (see (\ref{5-10})). Also, if $s^r>0$, the upstream structure is realised only partially. The combined solution (\ref{5-10}), (\ref{revrare}) is shown in Fig.~12. The weak discontinuities at the corners of the hydrodynamic rarefaction wave are resolved by small-amplitude dispersive wavetrains  which asymptotically decay with time.

\subsection{Downstream solution}

The dispersive resolution of the downstream hydraulic jump occurs between the states  $u^-=u^d<0,$ and $u^+=0$.  An inspection of the classification presented in Fig.~\ref{classpos} shows that for this range of $u^+, u^-$ the only possible solution is the normal cnoidal undular bore corresponding to Region 1 in the Riemann problem classification in Fig.~7.

The oscillatory structure of the bore is described by the elliptic solution (\ref{5-3}) with the parameters $u_j$, $j=1, \dots, 4$ expressed in terms of the Riemann invariants $r_i$, $i=1,2,3$  via relationships (\ref{8-4}). The spatiotemporal behaviour of the Riemann invariants is given by the Gurevich-Pitaevskii
similarity modulation solution  (\ref{9-3}) with constant Riemann invariants
\begin{equation}\label{9-8}
    r_1=u^d(1-\al u^d)<0,\quad r_3=0,
\end{equation}
so that expressions (\ref{8-4}) for the parameters $u_k$ (Region 1) reduce to
\begin{equation}\label{9-9a}
    \begin{split}
    u_1(r_2)&=\frac1{2\al}(1+2\al u^d-\sqrt{1-4\al r_2}),\\
    u_2(r_2)&=\frac1{2\al}(-1+2\al u^d+\sqrt{1-4\al r_2}),\\
    u_1(r_2)&=\frac1{2\al}(1-2\al u^d-\sqrt{1-4\al r_2}),\\
    u_1(r_2)&=\frac1{2\al}(3-2\al u^d+\sqrt{1-4\al r_2}),
    \end{split}
\end{equation}
and their plots are shown in Fig.~13a. The corresponding modulated solution for the downstream undular bore is shown in Fig.~13b.
The speed of the soliton edge is
\begin{equation}\label{9-9}
    s^-_d=\Delta-2u^d(1-\al u^d),
\end{equation}
and of the linear edge
\begin{equation}\label{9-10}
    s^+_d=\Delta-12u^d(1-\al u^d).
\end{equation}
\begin{figure}
\begin{center}
\includegraphics[width=6cm, clip]{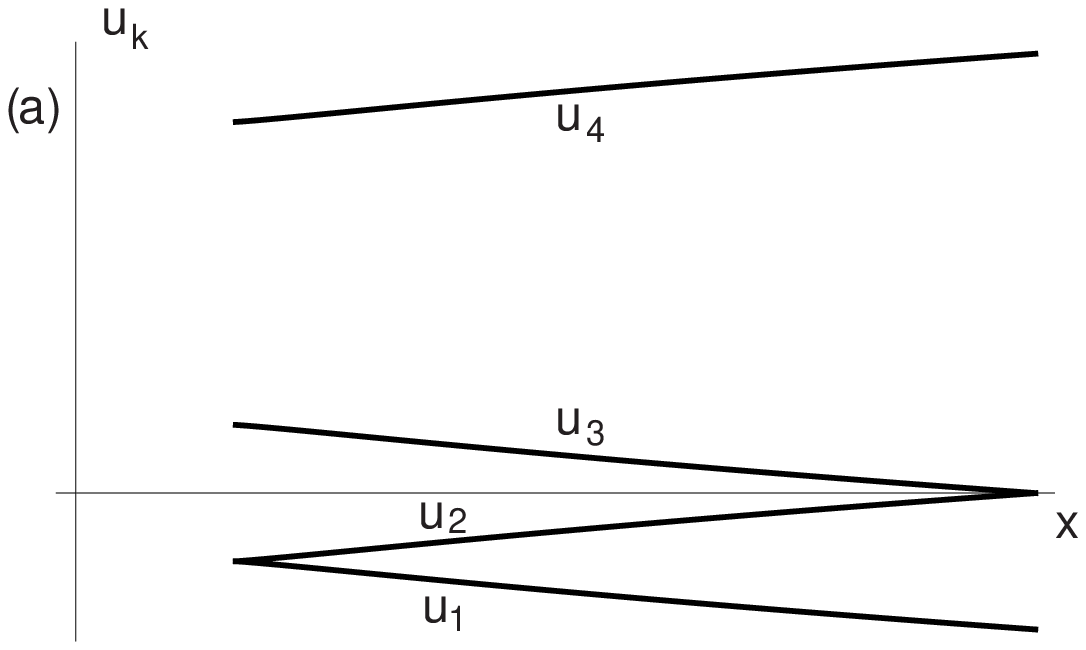} \qquad
\includegraphics[width=6cm, clip]{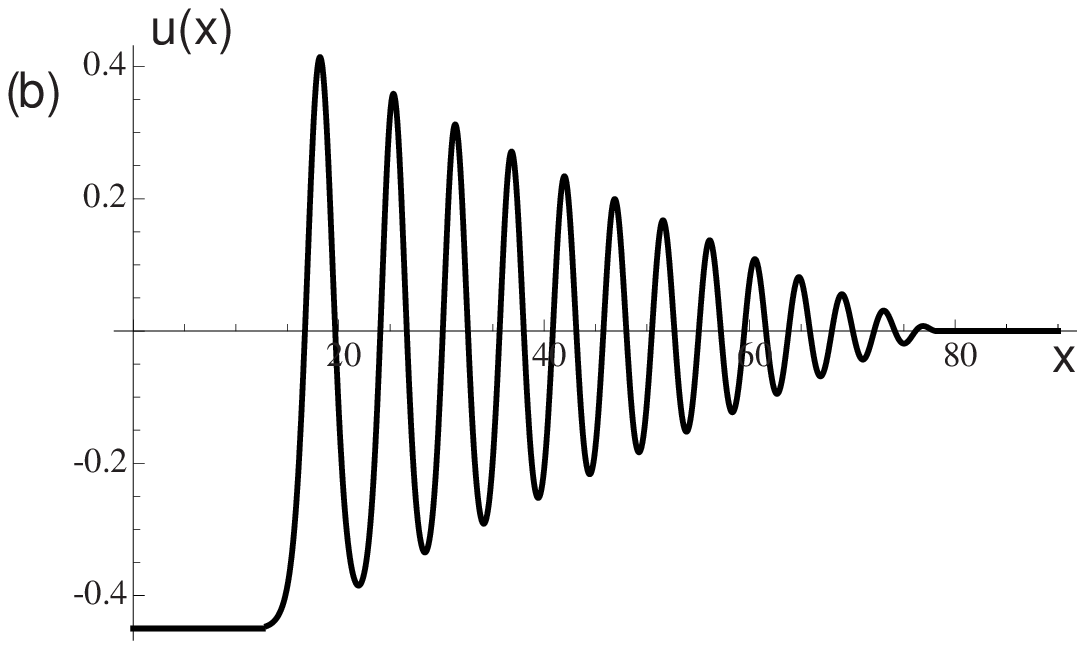}
\caption{Riemann problem solution to the unforced Gardner equation (\ref{1-1}) with $\alpha>0$ and the initial step parameters  $u^-=u^d<0$, $u^+=0$ (Region 1 in Fig.~7) corresponding to the generation of normal cnoidal undular bore downstream the obstacle.
(a) Typical behaviour of the modulation parameters $u_k$ on $x$ for  some fixed value of time $t>0$.
The point with $u_1=u_2=u^d$ corresponds to the soliton edge of the bore; the point with $u_2=u_3=0$ corresponds to the zero-amplitude (linear) edge.
(b) Plot of the analytical (modulation theory) solution to the unforced Gardner equation (\ref{1-1}) with $\al=1$, $\Delta=0$ and
the step-like initial condition (\ref{3-4b}) with $u^-=-0.45,$ $u^+=0.0$ at the time $t=10$.
}
\end{center}
\label{fig13}
\end{figure}
The downstream bore is attached to the topography if $s_-^d<0$.  Similar to the upstream bore, the condition $s_-^d=0$ yields the `critical' value of $u^d$, which, after the substitution in (\ref{3-7}) gives the expression of the curve $\Delta^d$ separating the regimes corresponding to the attached and detached downstream undular bore
\begin{equation}\label{und-5}
 \Delta^d: \ \    G_m(\Delta)=\frac1{2\al^2}\left\{\frac12\left(1-\sqrt{1-2\al\Delta}\right)^2-1+\al\Delta+
    \left(1-\frac{2\al\Delta}3\right)^{3/2}\right\}.
\end{equation}
For small $ \alpha |\Delta| \ll 1 $ equation (\ref{und-5}) is
approximated by the series expansion (cf. analogous expansion (\ref{u-6}) for the upstream bore)
\begin{equation}\label{u-61}
  G_m\alpha^2\cong \frac{1}3 ( \alpha \Delta)^2+\frac{7}{27}( \alpha \Delta)^3.
\end{equation}
Again, one should take the branch of the inverse function to (\ref{und-5}) which is located within the region $\Delta_-<\Delta<\Delta_1$. The curve (\ref{und-5}) is shown in Fig.~10 (see section 4.1).

One should note that, unlike the partial bores in the upstream resolution, the modulus $m$ in the partial downstream bore ranges in some interval $0<m<m^*$, where $m^* \le 1$ so that the partial undular bore downstream generally represents an extended ``rear'' part of the bore rather than the solitary wavetrain. This type of modulated solutions arises in the KdV initial boundary problem in the negative quarter plane and was studied in \cite{ms-2002}. Since the Gardner modulation theory for the considered region of parameters is  equivalent that for the KdV case we do not consider the details of the downstream partial bores here.

We remark that both curves (\ref{u-61}) and (\ref{u-6}) agree to leading order in $|\alpha \Delta| \ll 1$ with the equation of the (single) curve separating the regimes of the attached and detached upstream/downstream bores in the forced KdV equation theory \cite{gs-1986}, \cite{smyth-1987}. Note that in the forced KdV case, one of the bores is necessarily attached to the obstacle. Contrastingly, in the forced Gardner equation case, there is  the region $\Delta^d <\Delta < \Delta^u$ corresponding to the generation of two detached, fully realised undular bores.  A similar configuration with two complete bores occurs in fully nonlinear transcritical single-layer shallow-water flows described by the forced Su-Gardner equation  \cite{egs-2009}.

\subsection{Numerical simulations}

The previous  study \cite{gcc-2002} showed an excellent agreement of the transcritical hydraulic solution to the forced Gardner equation (\ref{1-1}) with the results of full numerical simulations for a broad range of the input parameters. On the other hand, in \cite{kamch-12} a detailed comparison of the analytical unsteady modulation solutions for the step resolution problem for the unforced Gardner equation with the numerical solutions of the same problem was carried out. The quantitative comparison was made for all regions in the classifications presented in Figs.~7 and 8 and a very good agreement with numerics was demonstrated.

Here we mostly concentrate on the verification of the qualitative predictions of our classification of the upstream and downstream unsteady wave patterns detailed in the  two previous subsections. As  is usually the case with the dispersive transcritical flow problems, the quantitative agreement is very good in the near-critical regime ($\Delta \approx 0$) and is getting worse when one moves towards the boundaries of the transcritical region of parameters
(see e.g. \cite{egs-2009} or \cite{legk-2009}). This is due to much longer times of the establishment of the hydraulic solution far from criticality resulting in a formal violation of the applicability conditions for the \cite{gs-1986} construction.

To verify the predictions of our analysis in Sections 4.1 and 4.2 we have performed three series of simulations
using the input parameters $\alpha$, $G_m$ and $\Delta$ from the three distinct subregions of the region
$\Delta_- < \Delta< \Delta_+'$ in the parametric flow-regime diagram in Fig.~4a.  In our numerical simulations
of the forced Gardner equation (\ref{1-1}) the standard pseudo-spectral method was used for treatment of the space
dependence together with the 4th order Runge-Kutta procedure for time stepping (see e.g. \cite{tref}).
The typical numerical results from all three series are presented in Figs. 14--17. As one can see, the numerical
simulations completely agree with the predictions of our classification.
\begin{figure}
\begin{center}
\includegraphics[width=6.2cm, clip]{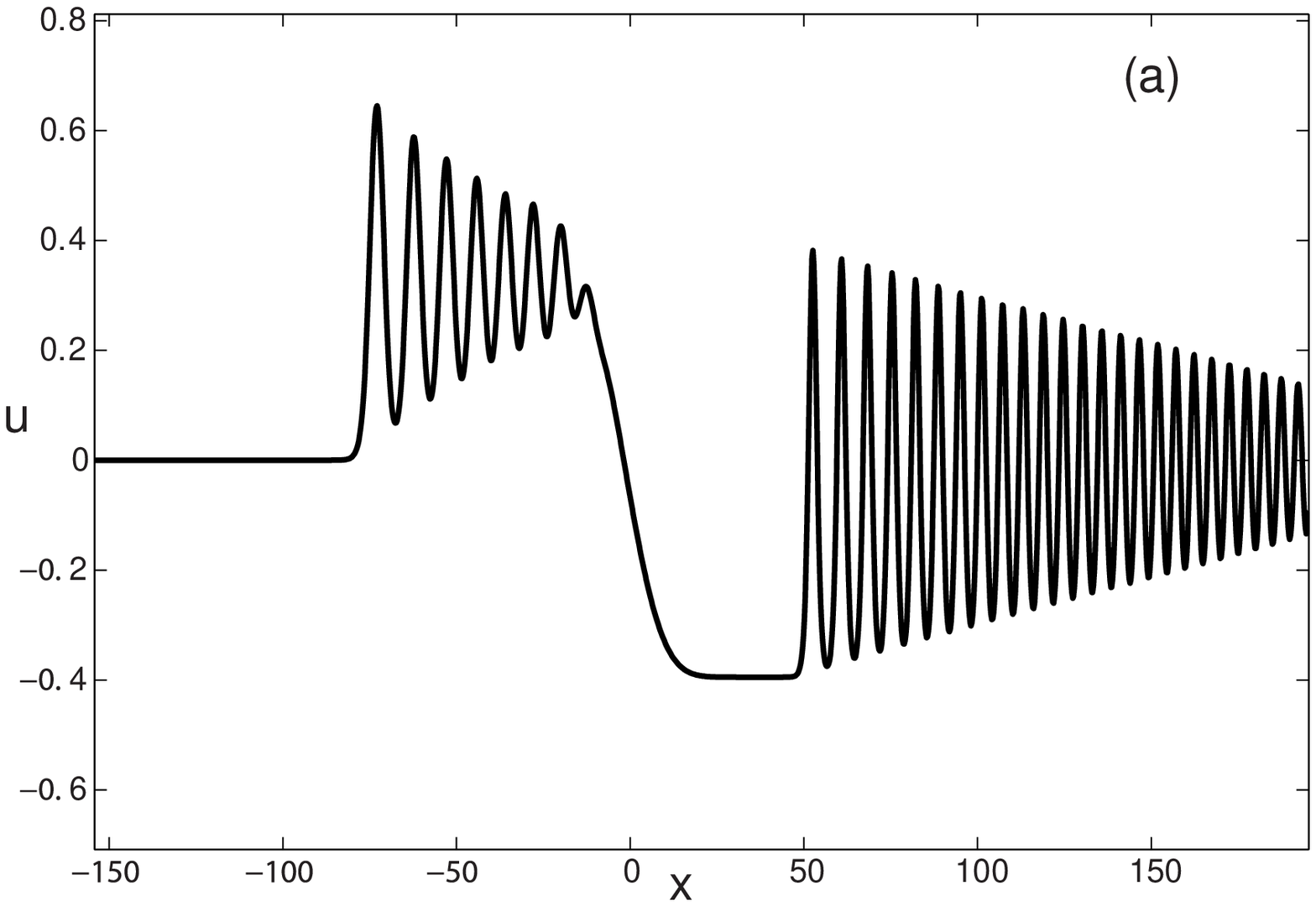} \quad
\includegraphics[width=6.2cm, clip]{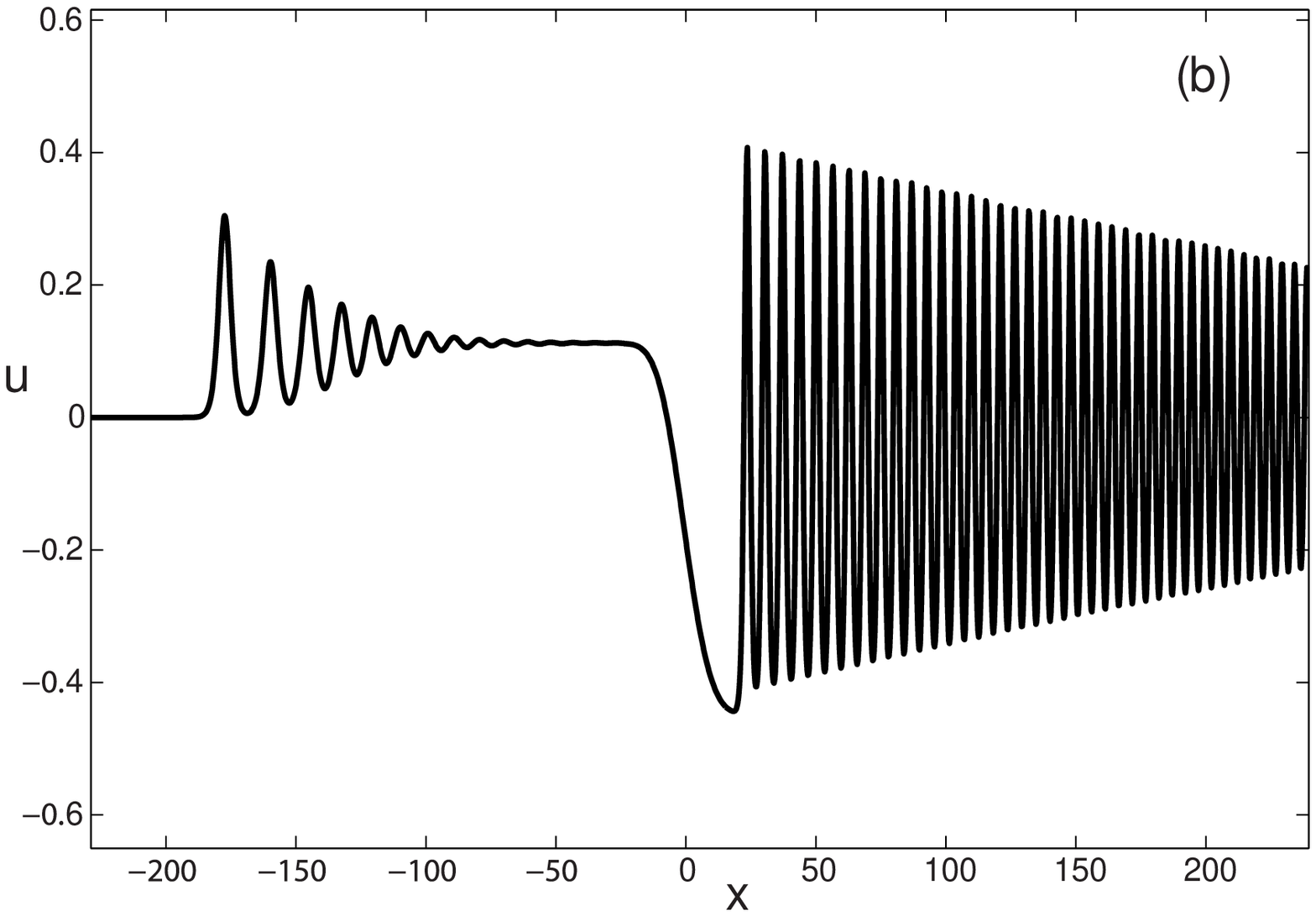}
\caption{Numerical solutions of the forced Gardner equation (\ref{1-1}) with $\alpha =1$ for different parts of the region $\Delta_-<\Delta<\Delta_1$ in the parametric flow-regime diagram in Fig. 4a (see also Fig.~10): (a) $\Delta_-< \Delta < \Delta^d$, ($\Delta =-0.44, G_m=0.43$, $t=55$): the upstream bore is attached, the downstream bore is detached; (b) $\Delta^u< \Delta < \Delta_1$, ($\Delta =-1.34, G_m=0.32$, $t=55$): the downstream bore is attached, the upstream bore is detached. In both cases the characteristic width of the topography forcing (\ref{1-3}) was taken $l=10$.
}
\end{center}
\label{fig14}
\end{figure}
\begin{figure}
\begin{center}
\includegraphics[width=8cm, clip]{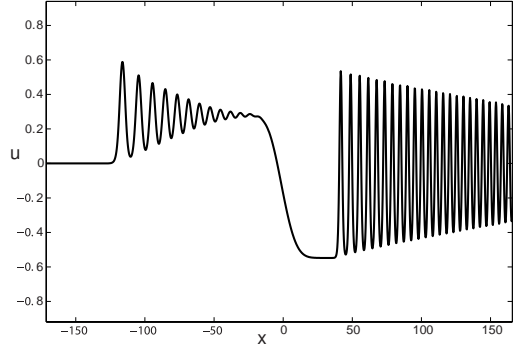}
\caption{Numerical solution of the forced Gardner equation (\ref{1-1}) with $\alpha =1$ for the region $\Delta^d< \Delta < \Delta^u$ of the input parameters (see Figs.~4a and 10): both undular bores are detached from the obstacle.  The parameters in the numerical solutions are $\Delta =-1.25, G_m=0.66$, $t=95$. The characteristic width of the topography forcing (\ref{1-3}) was taken $l=10$.
}
\end{center}
\label{fig15}
\end{figure}

\begin{figure}
\begin{center}
\includegraphics[width=7cm, clip]{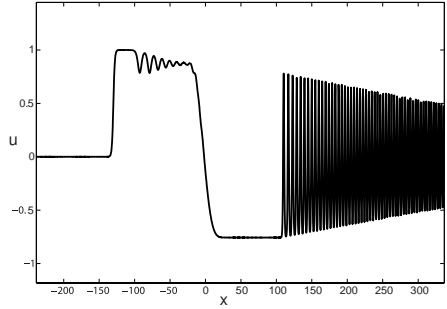}
\caption{Numerical solution of the forced Gardner equation (\ref{1-1}) with $\alpha =1$ for the region $\Delta_1< \Delta < \Delta_2$ of the input parameters (see Fig.~4a and 10). Upstream is an attached reversed undular bore connected to the equilibrium flow by a solibore (cf. the relevant analytical solution plot in Fig.~11b); downstream a detached normal undular bore is generated. The parameters in the numerical solution are: $\alpha=1$, $\Delta =-1, G_m=1.9, l=10$, $t=60$.
}
\end{center}
\label{fig16}
\end{figure}

\begin{figure}
\begin{center}
\includegraphics[width=12cm, clip]{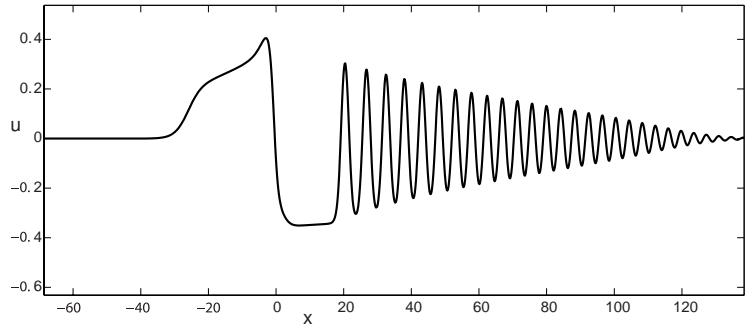}
\caption{Numerical solution of the forced Gardner equation (\ref{1-1}) with $\alpha =1$ for the region $\Delta_2< \Delta < \Delta'_+$ of the input parameters (see Fig. 4a). Upstream resolution pattern: a combination of a rarefaction wave  and a solibore (cf. the relevant analytical solution plot in Fig.~12); downstream -- a detached undular bore. The parameters in the numerical solution are: $\alpha=4.8$, $\Delta =-1, G_m=0.33$,  $l=4$, $t=20$.
}
\end{center}
\label{fig17}
\end{figure}

\bigskip
\section{Upstream and downstream closure, $\alpha<0$.}

An analysis of the dispersive resolution of the upstream and downstream  hydraulic jumps in the transcritical Gardner equation with  $\alpha<0$ is in many respects analogous to that performed in sections 4.1 and 4.2 for the case $\alpha>0$. However, due to  some distinctive features of the periodic travelling wave solutions of the Gardner equation with $\alpha<0$ resulting in non-strict hyperbolicity of the modulation system, the structure of the downstream and upstream waves generated in the transcritical Gardner flows  can drastically differ from that in the dispersive resolution patterns described in the previous section. The special role in the dispersive step resolutions in the Gardner equation (\ref{1-1}) with $\alpha<0$ is played (for a certain range of the initial step parameters) by the so-called trigonometric undular bores, which were first described in \cite{march2008} for the focusing mKdV equation, and which are, in some sense, the counterparts of the solibores generated in the `lock-exchange' flows of the Gardner equation with $\alpha>0$. The detailed analysis of the Gardner equation trigonometric bores can be found in \cite{kamch-12}.

\subsection{Upstream solution}

Similar to the case $\alpha>0$, in order to describe the upstream wave pattern we need to solve the unforced Gardner equation (\ref{1-1}) with  initial conditions (\ref{3-4b}) where $u^-=0$ and $u^+=u^u>0$. The dependence of $u^u$ on the definitive parameters $\Delta$, $G_m$ and $\alpha$ is given by the same equations (\ref{3-7}), (\ref{4-1}).
The value of $u^u$ within the region of existence of the localised transcritical hydraulic solution (see Fig.~4b) ranges now
from $(u^u)_{min}=0$ at the upper boundary $\Delta_+$  to some  $(u^u)_{\max}$ at the lower boundary $\Delta'_-$. The value $(u^u)_{\max}$ is found as the smaller positive root of equation (\ref{3-7}) with  $G_m(\Delta)$ defined by (\ref{3-3b}).

An inspection of the classification presented in Fig.~8 shows that within this range of $u^+=u^u$ (at fixed $u^-=0$) one can possibly have only one type of the upstream resolution -- the one corresponding to Region 4. The relevant structure is the KdV type ``normal'' cnoidal  undular bore with a bright soliton at the leading edge. Locally, this bore is described by the following periodic solution of the basic ODE (\ref{eq3}) (\cite{kamch-12}),
\begin{equation}\label{7-2}
    u=u_3+\frac{(u_4-u_3)\mathrm{cn}^2(\theta,m_2)}{1+
    \frac{u_4-u_3}{u_3-u_1}\mathrm{sn}^2(\theta,m_2)},
\end{equation}
where
\begin{equation}\label{6-4}
    \theta=\sqrt{|\al|(u_3-u_1)(u_4-u_2)}(x-x_0-Vt)/2,
\end{equation}
\begin{equation}\label{6-5}
    m=\frac{(u_4-u_3)(u_2-u_1)}{(u_4-u_2)(u_3-u_1)},
\end{equation}
and $V$ is given by (\ref{5-6}).

For Region 4, the expressions for the parameters $u_j$ of the periodic solution (\ref{7-2}) in terms of the Riemann invariants $r_1, r_2, r_3$ in the `universal' modulation solution (\ref{9-2}), (\ref{9-3}) are given by the following inverse  formulae of (\ref{ru}) (see
\cite{kamch-12}),
\begin{equation}\label{8-7}
    \begin{split}
    &u_1=\frac1{2\al}\left(1+\sqrt{1-4\al r_1}+\sqrt{1-4\al r_2}+\sqrt{1-4\al r_3}\right),\\
    &u_2=\frac1{2\al}\left(1-\sqrt{1-4\al r_1}-\sqrt{1-4\al r_2}+\sqrt{1-4\al r_3}\right),\\
    &u_3=\frac1{2\al}\left(1-\sqrt{1-4\al r_1}+\sqrt{1-4\al r_2}-\sqrt{1-4\al r_3}\right),\\
    &u_4=\frac1{2\al}\left(1+\sqrt{1-4\al r_1}-\sqrt{1-4\al r_2}-\sqrt{1-4\al r_3}\right).\\
    \end{split}
\end{equation}
The constant Riemann invariants (\ref{9-2}) are
\begin{equation}\label{10-1}
    r_1=0,\quad r_3=u^u (1-\al u^u),
\end{equation}
so that expressions (\ref{8-7}) assume the form
\begin{equation}\label{10-1a}
    \begin{split}
    u_1(r_2)&=\frac1{2\al}(3-2\al u^u+\sqrt{1-4\al r_2}),\\
    u_2(r_2)&=\frac1{2\al}(1-2\al u^u-\sqrt{1-4\al r_2}),\\
    u_3(r_2)&=\frac1{2\al}(-1+2\al u^u+\sqrt{1-4\al r_2}),\\
    u_4(r_2)&=\frac1{2\al}(1+2\al u^u-\sqrt{1-4\al r_2}).
    \end{split}
\end{equation}
The dependence of $r_2$ on $x,t$ is determined implicitly by the similarity modulation solution Eq.~(\ref{9-3}) so the modulations $u_j(x,t)$ (\ref{10-1a})
are completely defined.
When $r_2$ varies between $r_1$ and $r_3$ the parameters $u_j$ vary along the curves shown in Fig.~18a.
The corresponding plot of the undular bore constructed by substituting the modulation solution (\ref{10-1a}), (\ref{9-3}) in the travelling wave (\ref{7-2}) is shown in Fig.~18b.
\begin{figure}
\begin{center}
\includegraphics[width=6cm,clip]{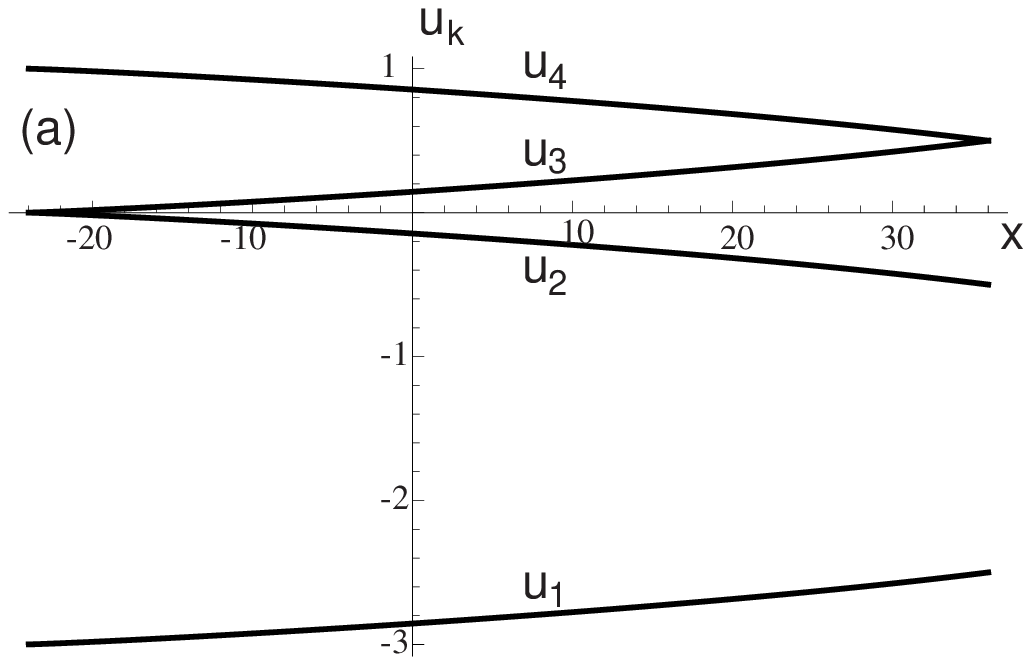} \quad
\includegraphics[width=6cm,clip]{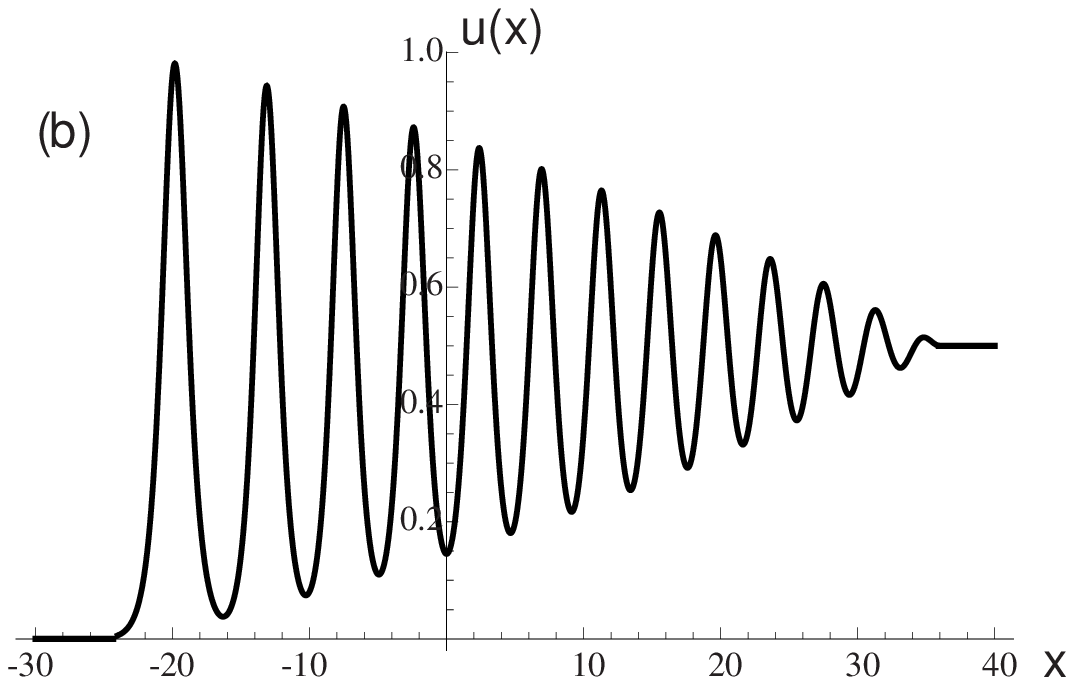}
\caption{Riemann problem solution to the  unforced Gardner equation (\ref{1-1}) with $\alpha <0$ and the initial step parameters
 $u^-=0$, $u^+=u^u>0$  corresponding to the upstream resolution via normal undular bore (Region 4 in Fig~8).
(a)  Typical behaviour of the modulation parameters $u_k$  in the  bore.
At the soliton edge $u_3=u_2=0$, at the linear zero-amplitude edge $u_3=u_4$;\
(b) The plot of the analytical (modulation theory) solution  to the unforced Gardner equation (\ref{1-1})
with $\al=-1$, $\Delta=0$ and the step-like initial condition (\ref{3-4b}) with $u^-=0,$ $u^+=0.5$ at the time $t=8$.
Only the upstream part, $x<0$, of the bore is realised in the transcritical flow.
}
\end{center}
\label{fig18}
\end{figure}
The speeds of the soliton and linear edges of the upstream undular bore  are
\begin{equation}\label{10-2}
    s^-_u=\Delta-4u^u(1-\al u^u) \quad \hbox{and} \quad s^+_u=\Delta+6u^u(1-\al u^u) \,
\end{equation}
respectively. Since $u^u>0$, $\alpha<0$ and $\Delta |\alpha|> -1$ (see Fig.~4b) it is not difficult to show that $s^+_u>0$ for all values of the input parameters $G_m, \Delta$ involved. Thus, for $\alpha<0$ the upstream undular bore can only be realised partially. As was mentioned earlier, the partial undular bore can often be approximated by a uniform train of solitary waves (see \cite{gs-1986}, \cite{smyth-1987} for the similar issue for the forced KdV equation). This approximation is expected to work particularly well for large values of the normalised topography amplitude $G_m \alpha^2$ leading to the ``cutoff'' modulus $m_0$ values sufficiently close to unity.

\smallskip

\subsection{Downstream solution}

To describe the downstream wave pattern we need to solve the unforced Gardner equation (\ref{1-1}) with
initial conditions (\ref{3-4b}) where
$u^-=u^d<0$ and $u^+=0$. The dependence of $u^d$ on the definitive parameters $\Delta$, $G_m$ is given by (\ref{3-7}), (\ref{4-1}).
The value of $u^d$ within the region of the existence of the localised transcritical hydraulic solution (see Fig.~4b) ranges
from $u^d=0$ at the upper boundary $\Delta_+$   to some  $(u^d)_{\min}<0$ at the lower boundary $\Delta'_+$. The value $(u^d)_{\min}$ is found as the negative root of equation (\ref{3-7}) (see Fig.~3b) with  $G_m(\Delta)$ defined by (\ref{3-3b}).

An inspection of the classification presented in Fig.~\ref{classneg} shows that, within this range of $u^-=u^d<0$ (at fixed $u^-=0$) one can possibly have three different types of behaviours corresponding to Regions 4, 3 and 2 respectively. The corresponding solutions are outlined below.

(a) $\Delta_1 < \Delta < \Delta_+$  This region of the parametric diagram in Fig.~4b corresponds to the initial steps (\ref{3-4b}) with $1/(2\al)<u^-<0,$ $u^+=0$, and, therefore, to Region 4 in the classification map in Fig.~8. Thus we have the classical KdV type ``bright''  cnoidal bore generated downstream the obstacle.
Similar to the upstream undular bore described in the previous subsection, the oscillatory structure of the undular bore for the considered range of the input parameters is described by the travelling wave solution (\ref{7-2})  with the relations (\ref{8-7}) between the modulation parameters $u_j, \ j=1,2,3,4$ and the  Riemann invariants $r_1,r_2,r_3$. The constant Riemann invariants in the  Gurevich-Pitaevskii solution (\ref{9-2}), (\ref{9-3}) now are
\begin{equation}\label{10-4}
    r_1=u^d(1-\al u^d),\quad r_3=0,
\end{equation}
which leads to the following set of expressions for $u_j$'s (cf. (\ref{10-1a}))
\begin{equation}\label{10-4a}
    \begin{split}
    u_1(r_2)&=\frac1{2\al}(3-2\al u^d+\sqrt{1-4\al r_2}),\\
    u_2(r_2)&=\frac1{2\al}(1+2\al u^d-\sqrt{1-4\al r_2}),\\
    u_3(r_2)&=\frac1{2\al}(-1+2\al u^d+\sqrt{1-4\al r_2}),\\
    u_4(r_2)&=\frac1{2\al}(1-2\al u^d-\sqrt{1-4\al r_2}).
    \end{split}
\end{equation}
\begin{figure}
\begin{center}
\includegraphics[width=6cm,clip]{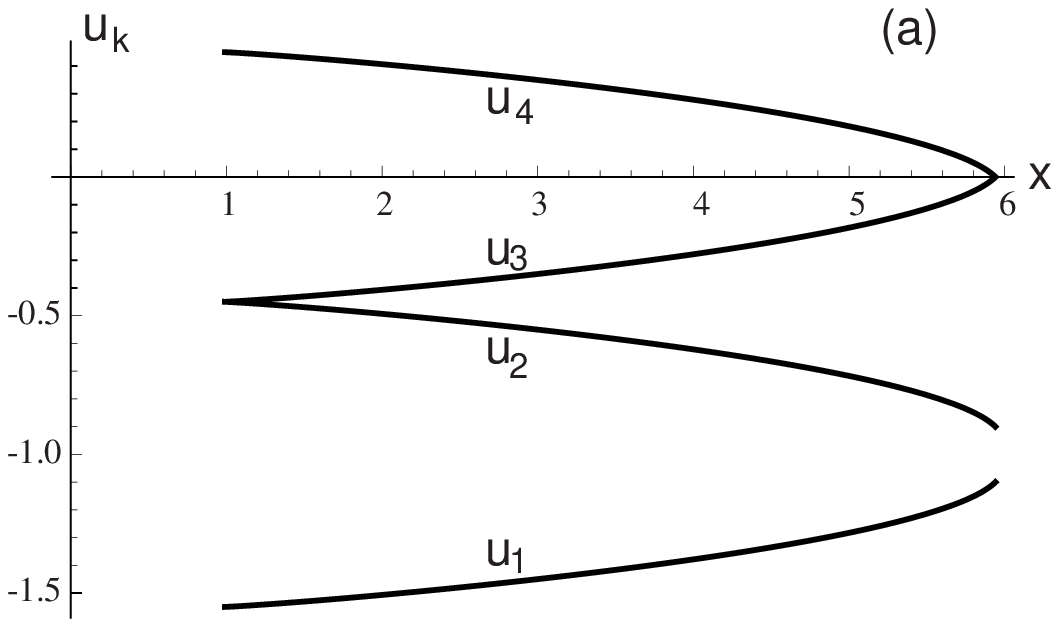} \quad
\includegraphics[width=6cm,clip]{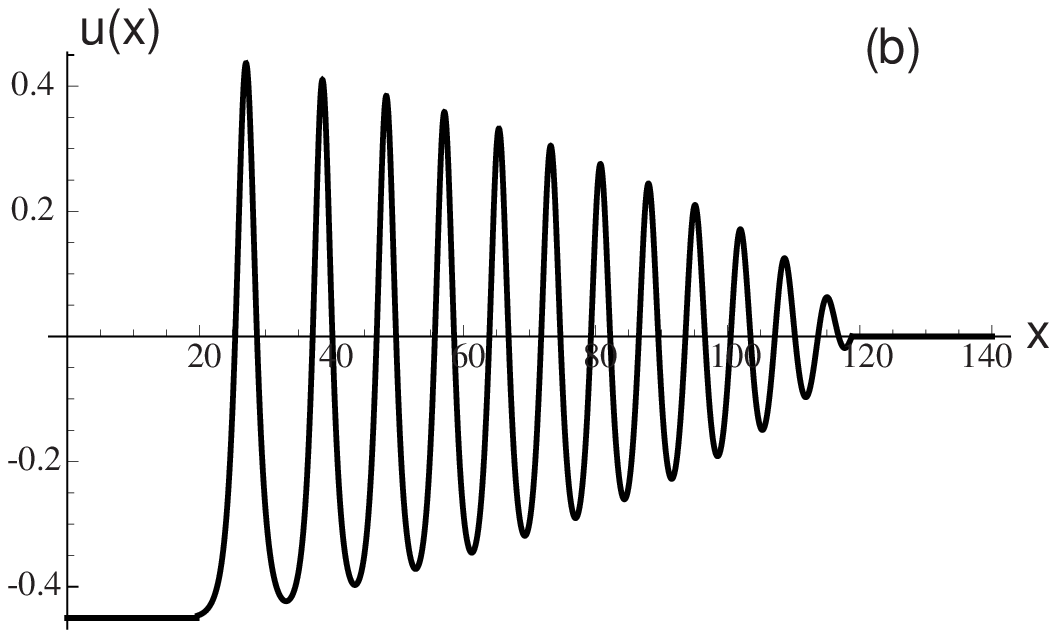}
\caption{Riemann problem solution to the unforced Gardner equation (\ref{1-1}) with $\alpha<0$ and the initial step parameters $u^+=0$, $u^-=u^d$, where
$1/(2\al)<u^d<0$, corresponding to the region $\Delta_1 < \Delta < \Delta_+$ in the transcritical flow-regime diagram in Fig.~4b (Region 4 in Fig. 8). (a) Typical behaviour of the modulation parameters
$u_k$ in the normal undular bore generated downstream the obstacle in the transcritical flow. At the soliton edge  $u_3=u_2=u^d$; at the linear edge $u_3=u_4=0$. (b) Plot of the analytical (modulation theory) solution to the Gardner equation with $\alpha=-1$,
$\Delta=0$ and the step-like initial condition (\ref{3-4b}) with $u^-=-0.45,$ $u^+=0.0$ at the time $t=40$.
 }
\end{center}\label{fig19}
\end{figure}
The speeds of the downstream undular bore soliton and  linear edges are
\begin{equation}\label{10-5}
    s^-_d=\Delta-2u^d(1-\al u^d) \quad \hbox{and} \quad  s^+_d=\Delta-12u^d(1-\al u^d)
\end{equation}
respectively. The soliton amplitude is $a=-2u^d$. The condition $s^-_d=0$ corresponds to the bore whose
soliton edge is attached to the obstacle. Expressing $u^d$ from this condition and substituting it into
equation (\ref{3-7}) we obtain the same equation (\ref{und-5}) obtained for the line $\Delta^d$ separating
the regimes of the attached (partial) and detached (fully realised) downstream undular bore for the case
$\alpha>0$. The only difference is that now, upon inversion of (\ref{und-5}), one chooses the upper branch
of the function $\Delta = \Delta^d (G_m)$. As a result, the partial downstream bore can only be realised
in a very small sector of parameters  with $\Delta <0$ in the flow-regime diagram in Fig.~4b. One should
also note that, unlike partial bores in the upstream resolution for $\alpha>0$ (see Section 4) the modulus
$m$ in the partial downstream bore ranges in some interval $0<m<m_0$, where $m_0 \le 1$. This type of modulated
solution for the KdV equation was studied in \cite{ms-2002}. Since the Gardner modulation theory for the
considered region of parameters is  equivalent that for the KdV case, we do not consider the details of
the downstream partial bores here.

\medskip

(b) $\Delta_2 < \Delta < \Delta_1$. This region of the parametric diagram in Fig.~4b corresponds
to the initial steps (\ref{3-4b}) with $1/\al<u^-<1/(2\al)$, $u^+=0$, and, therefore,
to Region 3 in the classification map in Fig.~8.

The wave structure generated downstream is now a composite undular bore, which consists of two parts.
The front (left) part is described by the cnoidal solution (\ref{7-2}) in which the modulus $m$ varies
from $m=1$ at the leading edge to $m=0$ at some point $x^*$, which, however is not the bore's trailing
edge as the amplitude $a$ does not vanish at this point. The structure then continues via the modulated
nonlinear trigonometric solution of the Gardner equation,
in which $u_2=u_1$ (i.e. $m=0$ -- see (\ref{6-5})) but $a = u_4-u_3 \ne 0$. This solution has the form
\begin{equation}\label{7-7}
    u=u_3+\frac{(u_4-u_3)\cos^2\theta}{1+
    \frac{u_4-u_3}{u_3-u_1}\mathrm{sin}^2\theta},
\end{equation}
where
\begin{equation}\label{7-8}
\begin{split}
    &\theta=\sqrt{|\al|(u_3-u_1)(u_4-u_1)}(x-x_0-Vt)/2,\\
    &V=\Delta-4u_1-\alpha(u_3u_4-3u_1^2).
    \end{split}
\end{equation}
The modulated trigonometric solution (\ref{7-7}) describes the rear (right) part of the downstream composite undular bore through the trailing edge where $a \to 0$. The details of the analytical description of the `trigonometric' undular bores in the Gardner equation can be found in \cite{kamch-12}; here we  apply the relevant solutions from \cite{kamch-12} to the particular configuration of the downstream resolution in the transcritical flow.

The modulation in the cnoidal part of the bore is described by the same expressions (\ref{10-4a}), where $r_2(x,t)$ varies
according to the Gurevich-Pitaevskii solution (\ref{9-2}), (\ref{9-3}) with the same as in (\ref{10-4}) constant Riemann invariants
\begin{equation}\label{tr-1}
    r_1=u^d(1-\al u^d),\quad r_3=0.
\end{equation}
The difference between the solution in the present case $1/\alpha <u^d<1/(2 \alpha)$ and the previous case (a)  $1/(2\al)<u^-<0,$ $u^+=0$ is that now one has to choose in (\ref{10-4a}) the other branch of $\sqrt{1-4r_2}$. As a result, the amplitude $a=u_4-u_3$ does not vanish at the edge with $r_1=r_2\,(m=0)$. Indeed, one can readily see that $a=8\alpha u^d>0$ at this point $x=x^*$, which coincides with position of the trailing edge in the Gurevich-Pitaevskii solution (\ref{9-3}).
Then, to get the required matching with $u=0$ downstream, one has to continue the solution beyond $x=x^*$, via the modulated nonlinear trigonometric wave (\ref{7-7}) along which $\left.v_2\right|_{m=0}=x/t$
that is
\begin{equation}\label{tr-2}
    r_1=r_2=r=\frac1{12}\left(\Delta-\frac{x}t\right).
\end{equation}
This solution extends up to $x=x^+=(\Delta-3/\al)t$ where $u_3=u_4$ ($a=0$). The variations of the modulation parameters $u_k$
 in the trigonometric part of the bore are given by
\begin{equation}\label{tr-3}
    u_1=u_2=\frac1{\al},\quad u_3=\frac1{\al}\sqrt{1-\frac{\al}3\left(\Delta-\frac{x}t\right)},
    \quad u_4=-\frac1{\al}\sqrt{1-\frac{\al}3\left(\Delta-\frac{x}t\right)}.
\end{equation}
\begin{figure}
\begin{center}
\includegraphics[width=6cm,clip]{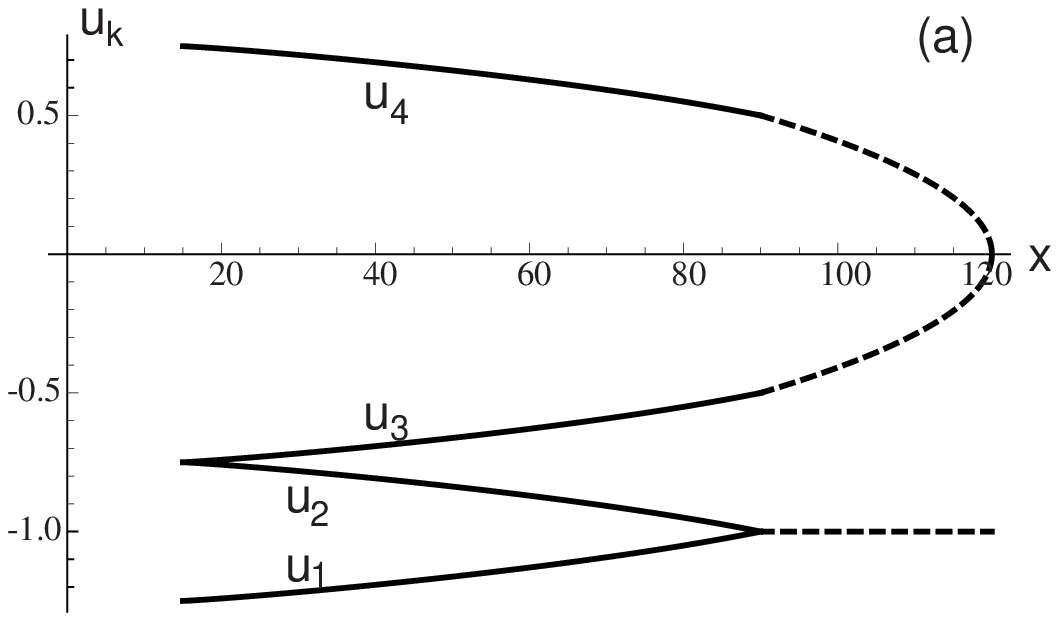} \qquad
\includegraphics[width=6cm,clip]{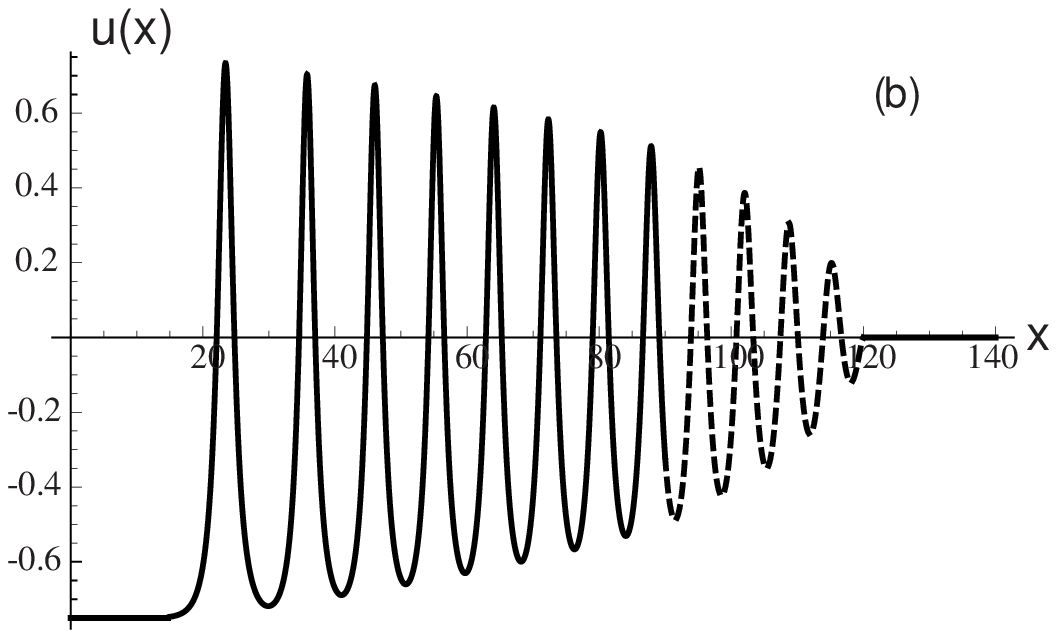}
\caption{Riemann problem solution to the unforced Gardner equation (\ref{1-1}) with $\alpha<0$ and $u^+=0$ and $u^-=u^d$, where $1/\al<u^d<1/(2\al)$ corresponding to the downstream resolution via a composite cnoidal-trigonometric undular bore in the transcritical flow with the parameters $\Delta, G_m$ located in the region $\Delta_2 < \Delta < \Delta_1$  of the parametric flow-regime diagram in Fig.~4b.   and Region 3  in the solution classification in Fig.~8.
(a) Typical behaviour of the modulation parameters $u_k$ in the  cnoidal-trigonometric undular bore. The point with $u_2=u_3=u^d$ corresponds to the leading, soliton edge;  the point with $u_3=u_4=0$ corresponds to the linear trailing edge. The boundary between the cnoidal and trigonometric part of the bore is located at the point where $u_2=u_1$. (b) Plot of the analytical (modulation theory) solution of the Gardner equation with $\al=-1$ and $\Delta=0$  and step-like initial conditions (\ref{3-4b}) with $u^-=-0.75,$ $u^+=0.0$ at the time $t=40$.
}
\end{center}
\label{fig20}
\end{figure}
The typical variations of $u_k$ as functions of $x$ in the composite cnoidal-trigonometric bore are shown in Fig.~20a and the corresponding oscillatory structure
is shown in Fig.~20b.  It is worth noticing that the presence of the trigonometric part of the bore leads to a slower
transition to the linear wave packet at the small amplitude edge compared with a pure cnoidal bore. The linear edge of the composite bore propagates with the velocity
\begin{equation}\label{10-6a}
    s^+=\Delta-\frac3{\al}.
\end{equation}
The boundary $x=x^*$ between the cnoidal and the trigonometric parts
propagates with the velocity
\begin{equation}\label{10-7}
    s^*=\Delta-12u^-(1-\al u^-)
\end{equation}
coinciding with the velocity of the trailing edge in the Gurevich-Pitaevskii solution (the second formula (\ref{10-5})).
The soliton edge propagates with the velocity $s^-$ given by the first formula (\ref{10-5}) and the lead soliton amplitude is

\medskip

c) $\Delta'_- < \Delta < \Delta_2$. This region of the parametric diagram in Fig.~4b corresponds to the initial steps (\ref{3-4b}) with $u^d_{\min}< u^-<1/\al$, $u^+=0$, where $u^d_{\min}$ is the smallest root of the equation (\ref{3-7}) with $\Delta=\Delta_-'$ defined by (\ref{3-3b}). This range of the initial steps corresponds to Region 2 in the classification chart in Fig.~8. Note that in this region $\Delta>0$.
The wave structure generated downstream is now a combination of a reversed rarefaction wave and a `normal'
trigonometric bore. This structure is a counterpart
of the combined solibore-rarefaction wave solution in the $\alpha>0$ case.
The trigonometric bore connects the equilibrium state $u=0$ at the right edge $x=s^+t$ and the `conjugate'
state $u=1/\alpha<0$ at the left edge $x=s^-t$ and is described by the periodic solution (\ref{7-7}), (\ref{7-8})
modulated according to (\ref{tr-3}). The edge speeds are
  \begin{equation}\label{tsp}
    s^- = \Delta \, , \qquad s^+= \Delta - \frac{3}{\alpha} \, .
  \end{equation}
Since in the region considered we have $\Delta>0$, the trigonometric bore is fully realised (cf. case (b) above). At the leading edge $x=s^-t$ the bore has an algebraic soliton, described by the limit of the trigonometric solution (\ref{7-7}) as $u_3 \to u_1$
\begin{equation}\label{eq13f}
\begin{split}
    &u=u_1+\frac{u_4-u_1}{1+|\alpha|(u_4-u_1)^2(x-Vt)^2/4},\\
    &V= \Delta - 6u_1(1-\alpha u_1).
    \end{split}
\end{equation}
\begin{figure}
\begin{center}
\includegraphics[width=8cm, clip]{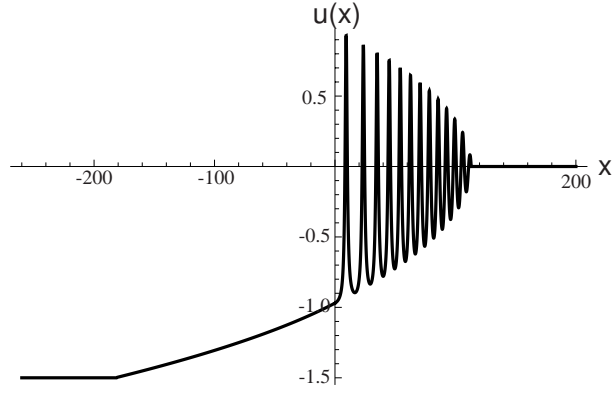}
\caption{Plot of the analytical (modulation theory) solution of the unforced Gardner equation (\ref{1-1}) with $\alpha<0$ and the initial parameters $u^+=0$, $u^-=u^d$, where  $u^d_{\min}< u^d<1/\al$, corresponding to the
downstream resolution of the transcritical flow with parameters $\Delta$, $G_m$ located in the region  $\Delta'_- < \Delta < \Delta_2$ of the parametric flow-regime diagram in Fig.~4b.  The resolution pattern consists of the reversed rarefaction wave and normal trigonometric bore joined together at the point with $u=1/\alpha$ (see Region 4 in the classification chart in Fig.~8).
The plot parameters are: $\alpha=-1$, $\Delta=0$, $u^-=-1.5$, $u^+=0$, $t=40$.}
\end{center}
\label{fig21}
\end{figure}
Using relationship (\ref{11-1}) and the fact that the pedestal of the algebraic soliton $u=u_1=1/\alpha$ we reduce (\ref{eq13f}) to the form
\begin{equation}\label{als}
u=\frac{1}{\alpha}\left(1-\frac{2}{1+\frac{4}{|\alpha|}(x-\Delta t)}\right) \, .
\end{equation}
The algebraic soliton (\ref{als}) is  joined to the downstream transcritical state $u=u^d$ right behind  the obstacle by the reversed rarefaction wave, which is asymptotically described by the following similarity solution of the dispersionless Gardner equation (\ref{1-1}) (cf. (\ref{revrare}))
  \begin{eqnarray}\label{revrare1}
u &=& u^d,  \quad \hbox{for}  \quad   x < s^l  t \, ,\nonumber \\
u &=& \frac1{2\al}\left(1+\sqrt{1+\frac{ 2\alpha } {3}\left(\frac{x}{t} - \Delta\right)}\right)  \quad \hbox{for} \quad  s^l t < x <  s^r t \, , \nonumber \\
 u &=& \frac1{\al} \quad \hbox{for} \quad x>s^r t   \, .
\end{eqnarray}
The speeds of the rarefaction wave edges are
\begin{equation}\label{rrw2}
s^l=\Delta - 6u^d(1-\alpha u^d), \qquad   s^r=\Delta\, ,
\end{equation}
so the speed of the right edge of the rarefaction wave coincides with the speed of the algebraic soliton in the
trigonometric bore implying that there is no additional constant states between the two structures.
Also it is not difficult to check that $0<s^l<s^r$ for the considered range of the input parameters so the
rarefaction wave never gets attached to the obstacle.
The above-described downstream composite solution is illustrated in Fig.~21.

\subsection{Numerical simulations}

In this section we present numerical solutions of the forced Gardner equation (\ref{1-1}) with $\alpha<0$ and the forcing term defined by formula (\ref{1-3}). Similar to the case $\alpha >0$ (see the numerical section 4.3), to verify the predictions of our analysis in Sections 5.1 and 5.2 we have performed three series of simulations using the input parameters $\alpha$, $G_m$ and $\Delta$ from the three distinct subregions of the region $\Delta_-' < \Delta< \Delta^+$ in the parametric flow-regime diagram in Fig.~4b. In Figs.~22--24 we present some of the typical upstream and downstream configurations illustrating our classification in the previous subsections. One can see that in all three plots the upstream bore is realised partially and attached to the obstacle as predicted by our analysis. Some irregularities in the structure of the leading part of the partial bore (or rather solitary wave train) in Figs.~22 and 24 is explained by the `history' of its generation. Namely, at the initial stage of the establishment of the steady transcritical hydraulic regime over the obstacle, a series of smaller solitary waves are generated. These  propagate downstream but eventually are overtaken by larger solitary waves continuously generated at the obstacle's location after the hydraulic transcritical transition has settled. The relative width of the irregular part of the bore decreases with time so for sufficiently  large $t$ can be safely neglected (see the long-time simulation in Fig. 23). In Fig.~23 one can also clearly see the ``wine-glass'' shape of the envelope of the downstream bore, which is the characteristic feature of the composite, cnoidal-trigonometric bore described in Section 5.2 and illustrated in the analytical plot in Fig.~20b. Finally, in Fig.~24 one can observe the downstream pattern consisting of an undular bore and a rarefaction wave, which agrees with the predicted theoretical pattern shown in  Fig.~21.
\begin{figure}
\begin{center}
\includegraphics[width=10cm, clip]{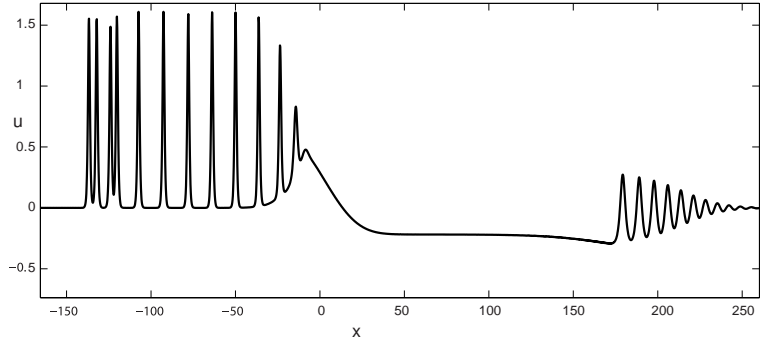} \qquad
\caption{Numerical solutions of the forced Gardner equation (\ref{1-1}) with $\alpha = - 1$ for  $\Delta_1<\Delta<\Delta_+$ (see the parametric flow-regime diagram in Fig. 4b). The generated waves are: the partial cnoidal undular bore (solitary wave train) upstream and fully realised cnoidal bore downstream. The plot parameters are: $\Delta =1.49, G_m=0.44, l=20, t=85$.
}
\end{center}
\label{fig22}
\end{figure}

\begin{figure}
\begin{center}
\includegraphics[width=12.5cm, clip]{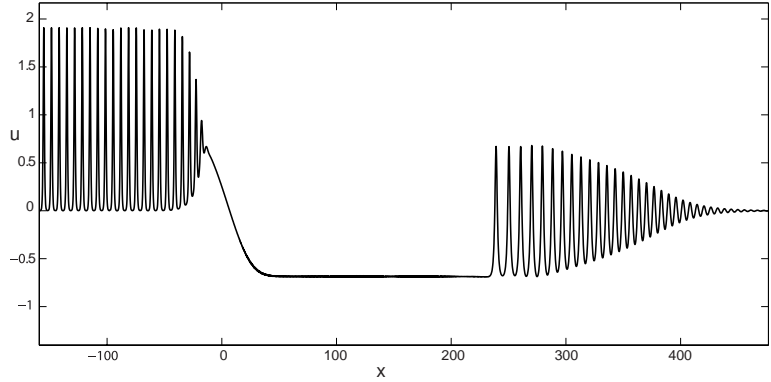}
\caption{Numerical solution of the forced Gardner equation (\ref{1-1}) with $\alpha =-1$ for $\Delta_+< \Delta < \Delta_2$ (see the parametric flow-regime diagram in Fig. 4b). The generated waves are: the partial cnoidal undular bore (solitary wave train) upstream and the composite cnoidal-trigonometric undular bore downstream (cf. the analytical curve in Fig.~20). The plot parameters are: $\Delta =1.89, G_m=2.31, l=20, t=85$.
}
\end{center}
\label{fig23}
\end{figure}

\begin{figure}
\begin{center}
\includegraphics[width=11cm, clip]{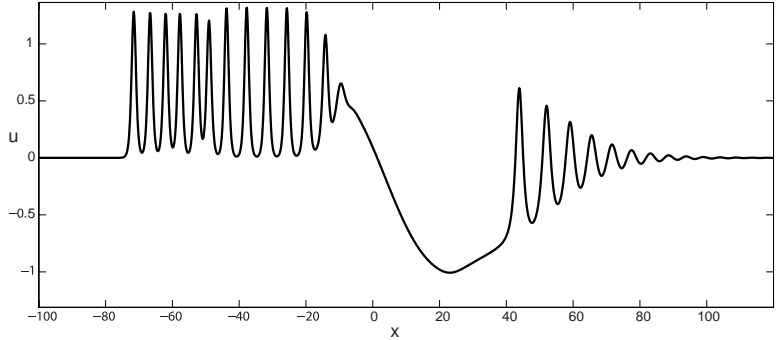}
\caption{Numerical solutions of the forced Gardner equation (\ref{1-1}) with $\alpha = - 1$ for the region  $\Delta_-'<\Delta<\Delta_2$ of the parametric flow-regime diagram in Fig. 4b. The generated waves are: the  partial cnoidal undular bore (solitary wave train) upstream and the composite pattern consisting of the normal trigonometric undular bore and the reversed rarefaction wave downstream (cf. analytical curve in Fig.~21). The plot parameters are: $\Delta =0.595, G_m=1.64, l=10, t=20$.
}
\end{center}
\label{fig24}
\end{figure}

\section{Conclusion}

The classification of the transcritical (resonant)  flows of a stratified fluid over topography is constructed in the framework of the Gardner (extended KdV) equation (\ref{1-1}) for a range of flow parameters, when the obstacle supports a steady hydraulic transition from the subcritical flow upstream to the supercritical flow downstream. We consider both possible signs for the cubic nonlinear term $\alpha$ in the Gardner equation corresponding to different fluid density stratification profiles. This problem has been considered in a number of earlier studies using various approaches, but mostly numerically with no full analytical description available.

The main contribution of the present paper is the development of the missing analytical theory of the transcritical Gardner flows in the framework of the \cite{gs-1986} approach, where the asymptotic description of the the flow is constructed using a combination of the hydraulic approximation over the obstacle's location and the modulation description of the unsteady waves generated upstream and downstream. The extension of the GS theory, which was originally developed for the forced KdV equation, to the forced Gardner equation is quite nontrivial and has become possible due to the recent development in \cite{kamch-12} of the full modulation theory for the (unforced) Gardner equation.

We have shown that, along with the regular cnoidal undular bores  occurring in the analogous problem for the single-layer flow modeled by the forced KdV equation and its non-integrable extensions (see \cite{egs-2009}), the forced Gardner equation supports a diverse family of upstream and downstream wave structures which are sharply different from those arising in the forced KdV equation dynamics. These new structures include combinations of solibores and rarefaction waves, `reversed' undular bores  and the so-called `trigonometric' undular bores. The classification of the transcritical flow  regimes is made in terms of two definitive parameters: the obstacle's height $G_m$ and the deviation $\Delta$ of the incident flow velocity from the long wave phase speed. This is done using a combination of the detailed analysis of the hydraulic solution of the forced Gardner equation with the dispersive term neglected and the relevant parts of the Riemann problem solution of the full unforced Gardner equation available from \cite{kamch-12}. The latter work has enabled the analytical description of the  unsteady wave structures generated upstream and downstream of the obstacle. The predictions of the constructed analytical theory are confirmed by direct numerical computations of the forced Gardner equation.

Apart from being fundamentally important as the basic analytical model for the weakly nonlinear transcritical flows in stratified fluids, the developed theory should provide a guidance for similar classifications
for the more advanced, fully nonlinear models of the transcritical internal flows past topography, for example,  the forced Myatta-Choi-Camassa equations for the two-layer flows past topography, which would generalise the theory of the fully nonlinear shallow-water single layer transcritical flows constructed in \cite{egs-2009}.

\subsection*{Acknowledgements}
We acknowledge helpful discussions with Mark Hoefer and Eduardo Khamis.
AMK thanks National Taiwan University and Taida Institute for Mathematical Sciences,
where this work was started, for kind hospitality.

\bibliographystyle{jfm}

\bibliography{gardner-topography}

\begin{thebibliography}{29}
\expandafter\ifx\csname natexlab\endcsname\relax\def\natexlab#1{#1}\fi

\bibitem[Akylas(1984)]{akylas84}
{\sc Akylas, T.~R.} 1984 On the excitation of long nonlinear water waves by a
  moving pressure distribution. {\em J. Fluid Mech.\/} {\bf 141}, 455--466.

\bibitem[Apel {\em et~al.\/}(2007)Apel, Ostrovsky \& Lynch]{aosl-2007}
{\sc Apel, J.~R., Ostrovsky, L.~A. \& Lynch, J.~F.} 2007 Internal solitons in
  the ocean and their effect on underwater sound. {\em J. Acoust. Soc. Am.\/}
  {\bf 121}, 695--722.

\bibitem[Baines(1984)]{baines84}
{\sc Baines, P.~G.} 1984 A unified description of two-layer flow over
  topography. {\em J. Fluid Mechanics\/} {\bf 146}, 127--167.

\bibitem[Baines(1995)]{baines}
{\sc Baines, P.~G.} 1995 {\em Topographic effects in stratified flows\/}.
  Cambridge University Press, Cambridge.

\bibitem[Cole(1985)]{cole85}
{\sc Cole, S.~L.} 1985 Transient waves produced by flow past a bump. {\em Wave
  Motion\/} {\bf 7}, 579--587.

\bibitem[El {\em et~al.\/}(2009)El, Grimshaw \& Smyth]{egs-2009}
{\sc El, G.~A., Grimshaw, R. H.~J. \& Smyth, N.~F.} 2009 Transcritical
  shallow-water flow past topography: finite-amplitude theory. {\em J. Fluid
  Mechanics\/} {\bf 640}, 187--214.

\bibitem[Esler \& Pierce(2011)]{ep-11}
{\sc Esler, J.~G. \& Pierce, J.~D.} 2011 Dispersive dam-break and lock-exchange
  flows in a two-layer fluid. {\em J. Fluid Mechanics\/} {\bf 667}, 555--585.

\bibitem[Fornberg \& Whitham(1978)]{fw78}
{\sc Fornberg, B. \& Whitham, G.~B.} 1978 A numerical and theoretical study of
  certain nonlinear wave phenomena. {\em Philosophical Transactions of the
  Royal Society A\/} {\bf 289}, 373--404.

\bibitem[Grimshaw(2001)]{grimshaw-2001}
{\sc Grimshaw, R.} 2001 {\em Environmental Stratified Flows\/}. Kluwer
  Academic, Boston.

\bibitem[Grimshaw {\em et~al.\/}(2002)Grimshaw, Chan \& Chow]{gcc-2002}
{\sc Grimshaw, R. H.~J., Chan, K.~H. \& Chow, K.~W.} 2002 Transcritical flow of
  a stratified fluid: the forced extended \uppercase{K}orteweg -- de
  \uppercase{V}ries model. {\em Phys. Fluids\/} {\bf 14}, 755--774.

\bibitem[Grimshaw \& Smyth(1986)]{gs-1986}
{\sc Grimshaw, R. H.~J. \& Smyth, N.~F.} 1986 Resonant flow of a stratified
  fluid over topography. {\em J. Fluid Mechanics\/} {\bf 697}, 237--272.

\bibitem[Gurevich \& Pitaevskii(1974)]{GP74}
{\sc Gurevich, A.~V. \& Pitaevskii, L.~P.} 1974 Nonstationary structure of a
  collisionless shock wave. {\em Sov. Phys. JETP\/} {\bf 38}, 291--297.

\bibitem[Helfrich \& Melville(2006)]{hm-2006}
{\sc Helfrich, K.~R. \& Melville, W.~K.} 2006 Long nonlinear internal waves.
  {\em Ann. Rev. Fluid Mech\/} {\bf 38}, 395--425.

\bibitem[Kakutani \& Yamasaki(1978)]{ky-1978}
{\sc Kakutani, T. \& Yamasaki, N.} 1978 Solitary waves on a two-layer fluid.
  {\em J. Phys. Soc. Japan.\/} {\bf 45}, 674--679.

\bibitem[Kamchatnov {\em et~al.\/}(2012)Kamchatnov, Kuo, Lin, Horng, Gou,
  Clift, El \& Grimshaw]{kamch-12}
{\sc Kamchatnov, A.~M., Kuo, Y.-H., Lin, T.-C., Horng, T.-L., Gou, S.-C.,
  Clift, R., El, G.~A. \& Grimshaw, R. H.~J.} 2012 Undular bore theory for the
  \uppercase{G}ardner equation. {\em Phys. Rev. E\/} {\bf 86}, 036605.

\bibitem[Kodama {\em et~al.\/}(2008)Kodama, Pierce \& Tian]{kod2008}
{\sc Kodama, Y., Pierce, V.~U. \& Tian, F.-R.} 2008 On the \uppercase{W}hitham
  equations for the defocusing complex modified \uppercase{K}d\uppercase{V}
  equation. {\em SIAM J. Math. Anal.\/} {\bf 41}, 26--58.

\bibitem[LeFloch(2002)]{leFloch}
{\sc LeFloch, P.~G.} 2002 {\em Hyperbolic systems of conservation laws\/}.
  Birkhauser.

\bibitem[Leszczyszyn {\em et~al.\/}(2009)Leszczyszyn, El, Gladush \&
  Kamchatnov]{legk-2009}
{\sc Leszczyszyn, A.~M., El, G.~A., Gladush, Yu.~G. \& Kamchatnov, A.~M.} 2009
  Transcritical flow of a \uppercase{B}ose-\uppercase{E}instein condensate
  through a penetrable barrier. {\em Phys. Rev. A\/} {\bf 79}, 063608.

\bibitem[Madsen \& Hansen(2012)]{madsen-2012}
{\sc Madsen, P.~A. \& Hansen, A.~B.} 2012 Transient waves generated by a moving
  bottom obstacle: a new near-field solution. {\em J. Fluid Mechanics\/} {\bf
  169}, 429--464,.

\bibitem[Marchant(2008)]{march2008}
{\sc Marchant, T.~R.} 2008 Undular bores and the initial-boundary value problem
  for the modified \uppercase{K}orteweg -- de \uppercase{V}ries equation. {\em
  Wave Motion\/} {\bf 45,}, 540Ð555.

\bibitem[Marchant \& Smyth(1990)]{ms-1990}
{\sc Marchant, T.~R. \& Smyth, N.~F.} 1990 The extended \uppercase{K}orteweg --
  de \uppercase{V}ries equation and the resonant flow of a fluid over
  topography. {\em J. Fluid Mechanics\/} {\bf 221}, 263--287.

\bibitem[Marchant \& Smyth(2002)]{ms-2002}
{\sc Marchant, T.~R. \& Smyth, N.~F.} 2002 The initial boundary problem for the
  \uppercase{K}orteweg -- de \uppercase{V}ries equation on the negative
  quarter-plane. {\em J. Fluid Mechanics\/} {\bf 458}~(2020), 857--871.

\bibitem[Melville \& Helfrich(1987)]{mh-1987}
{\sc Melville, W.~K. \& Helfrich, K.~R.} 1987 Transcritical two-layer flow over
  topography. {\em J. Fluid Mechanics\/} {\bf 178}, 31--52.

\bibitem[Smyth(1987)]{smyth-1987}
{\sc Smyth, N.} 1987 Modulation theory solution for resonant flow over
  topography. {\em Proc. Roy. Soc. Lond. A.\/} {\bf 409}, 79--97.

\bibitem[Trefethen(2000)]{tref}
{\sc Trefethen, L.~N.} 2000 {\em Spectral Methods in MATLAB\/}. Society for
  Industrial and Applied Mathematics, Philadelphia.

\bibitem[Wan {\em et~al.\/}(2010)Wan, Muenzel \& Fleischer]{fleischer}
{\sc Wan, W., Muenzel, S. \& Fleischer, J.~W.} 2010 Wave tunneling and
  hysteresis in nonlinear junctions. {\em Phys. Rev. Lett.\/} {\bf 104},
  073903.

\bibitem[White \& Helfrich(2012)]{wh2012}
{\sc White, B.~L. \& Helfrich, K.~R.} 2012 A general description of a gravity
  current front propagating in a two-layer stratified fluid. {\em J. Fluid
  Mechanics\/} {\bf 711}, 545--575.

\bibitem[Whitham(1965)]{whitham1}
{\sc Whitham, G.~B.} 1965 Non-linear dispersive waves. {\em Proc. Roy. Soc.
  A\/} {\bf 283}, 238--261.

\bibitem[Whitham(1974)]{whitham2}
{\sc Whitham, G.~B.} 1974 {\em Linear and Nonlinear Waves\/}.
  Wiley--Interscience, New York.

\end{thebibliography}

\end{document}